\documentclass[12pt,a4paper,dvips]{article}
\usepackage{a4p}
\usepackage{rotating}
\usepackage{cite}
\usepackage{graphicx}
\usepackage{epsfig}
\usepackage{physics}
\usepackage{l3_titlePH,ifthen}

\journalname{Phys. Lett. B}

\date{March 4, 2005}

\preprint{2005-011}

\newlength{\capindent}
\newlength{\capwidth}
\newlength{\figwidth}
\newcommand{\icaption}[2][!*!,!]{\hspace*{\capindent}%
  \begin{minipage}{\capwidth}
    \ifthenelse{\equal{#1}{!*!,!}}%
      {\caption{#2}}%
      {\caption[#1]{#2}}
  \end{minipage}}

\begin{document}
%

\begin{titlepage}

\title{\boldmath  Compton Scattering of\\ Quasi-Real Virtual Photons at LEP}

\author{L3 Collaboration}

\begin{abstract}

Compton scattering of quasi-real virtual photons, $\rm\gamma e^{\pm}
\rightarrow \gamma e^{\pm}$, is studied with 0.6~fb$^{-1}$ of data
collected by the L3 detector at the LEP $\epem$ collider at
centre-of-mass energies $\sqrt{s}=189-209\GeV$. About 4500 events
produced by the interaction of virtual photons emitted by $\rm e^\pm$ of
one beam with  $\rm e^\mp$  of the opposite beam are collected for
effective centre-of-mass energies of the photon-electron and
photon-positron systems in the range from $\sqrt{s'}= 35\GeV$ up to
$\sqrt{s'}=175\GeV$, the highest energy at which Compton scattering
was ever probed. The cross sections of the $\rm\gamma e^{\pm}
\rightarrow \gamma e^{\pm}$ process as a function of $\sqrt{s'}$ and
of the rest-frame scattering angle are measured, combined with
previous L3 measurements down to $\sqrt{s'}\simeq 20\GeV$, and found
to agree with the QED expectations.

\end{abstract}

\submitted
\end{titlepage}

%
\section{Introduction}
%

The CERN LEP $\epem$ collider allowed high-energy tests of high-order
QED through the detection of events with multiple hard-photon
production~\cite{LEPgammagamma} and the study of lepton pairs
produced by two-photon interactions~\cite{LEPleptons}.

\begin{figure}[p]
  \begin{center}
    \begin{tabular}{c}
      \mbox{\epsfig{figure=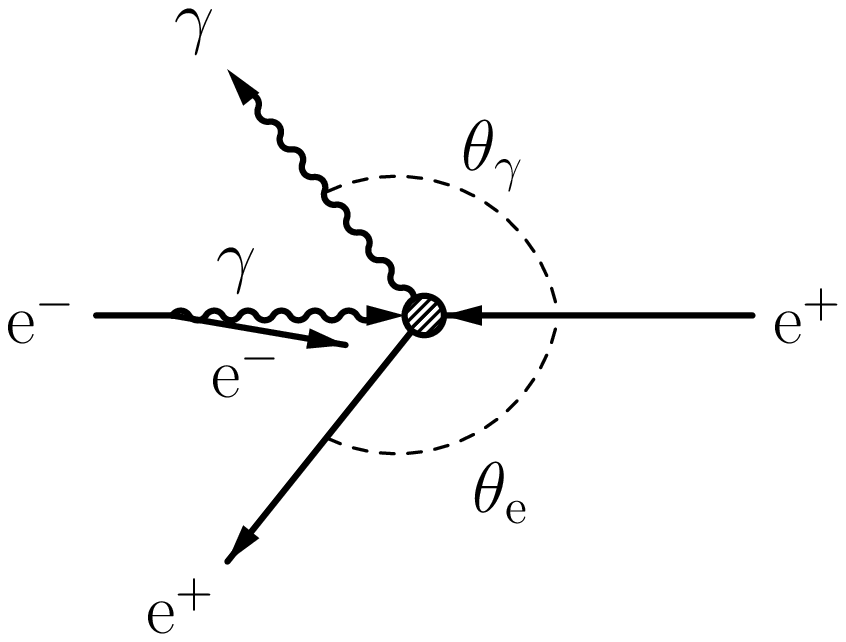,width=0.5\textwidth}}\\
      \hspace{1ex}\\
      \mbox{\epsfig{figure=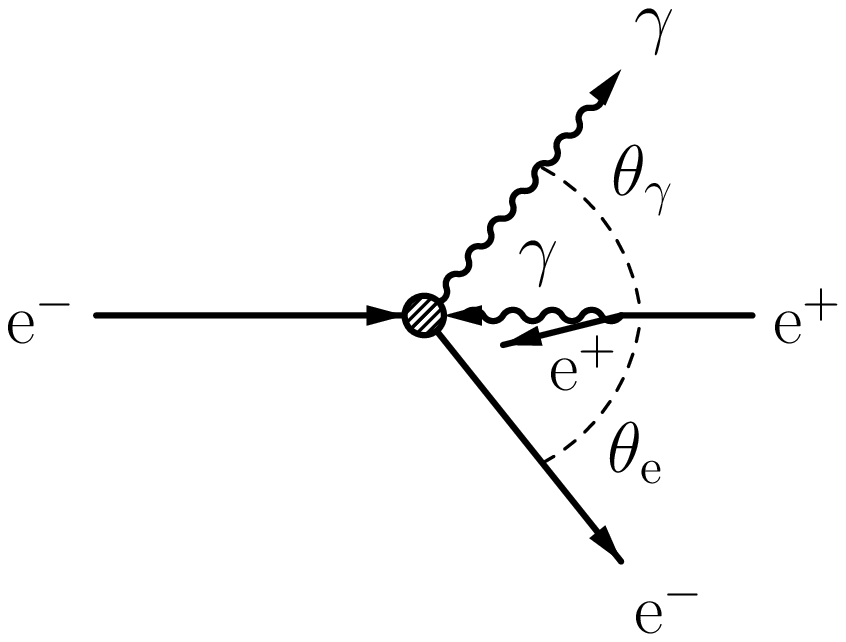,width=0.5\textwidth}}\\
      \hspace{1ex}\\
      \mbox{\epsfig{figure=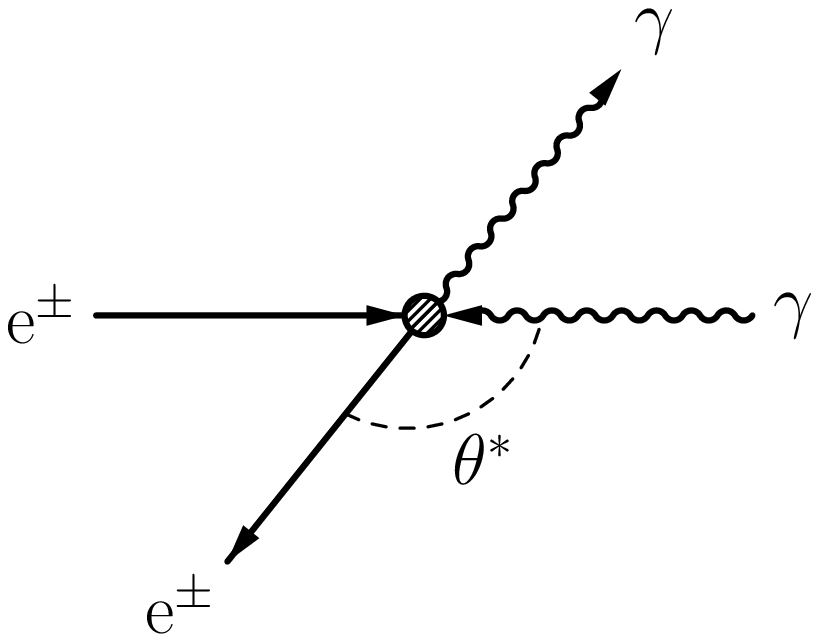,width=0.5\textwidth}}\\
    \end{tabular}
    \icaption[]{\label{fig:1} Schematic view of quasi-real Compton
    scattering in $\epem$ collisions in the laboratory system, higher plots, and in the
    $\gamma\rm e^{\pm}$ centre-of-mass frame, lowest plot.}
  \end{center}
\end{figure}

A unique test of QED at $\epem$ colliders is the study of Compton
scattering, $\rm\gamma e^{\pm} \rightarrow \gamma e^{\pm}$. In this
process, quasi-real virtual photons\footnote{A ``quasi-real'' virtual
photon is one whose virtuality, $Q^2=-m_\gamma^2$, is much smaller than all
relevant scales of the process and is therefore kinematically
equivalent to a real photon with $Q^2=0$.} emitted by one of the
incoming beams interact with the electrons\footnote{In this Letter,
the term ``electron'' is in general used to refer to both electrons
and positrons, unless specified otherwise.} of the other beam. This
process, $\epem\ra\epem\gamma$, is sketched in Figure~\ref{fig:1}. The
electron which radiates the quasi-real photon is scattered at a very
small angle and escapes detection along the beam pipe. The signature
of such a process is a photon and an electron in the detector, with a
large amount of missing momentum directed along the beam line. The
$\rm \gamma e^\pm \ra \gamma e^\pm$ process is characterised by the
effective centre-of-mass energy of the photon-electron collision,
$\sqrt{s'}$, and by the scattering angle of the electron in the
photon-electron centre-of-mass system, $\theta^*$, shown in
Figure~\ref{fig:1}. At the lowest order, the differential cross
section as a function of $\cos\theta^*$ is~\cite{HalzenMartin}:
\begin{equation}
	{\frac{\rm d\sigma}{\rm d\cos \theta^*} = \frac{\alpha^2
	\pi}{s'}\Bigg(\frac{1+\cos \theta^*}{2} + \frac{2}{1+\cos
	\theta^*}\Bigg)},
\label{eq:1}
\end{equation}
where $\alpha$ is the fine-structure constant. In the angular range
$|\cos \theta^*|<0.8$, this corresponds to a cross section of about
200~pb at $\sqrt{s'}=40\GeV$ and of about 20~pb at
$\sqrt{s'}=130\GeV$.

Quasi-real Compton scattering in $\epem$ colliders was first discussed
more than three decades ago~\cite{Carimalo} and observed at the
$900\MeV$ ACO storage ring in Orsay~\cite{Cosme}. The L3 Collaboration
studied this process at $\epem$ centre-of-mass energies
$\sqrt{s}=91-183\GeV$, covering with high statistics the effective
centre-of-mass energy range $\sqrt{s'}=20-100\GeV$~\cite{L3Old}. This Letter
presents the extension of this analysis to the high-luminosity and
high-energy data sample collected at LEP at $\sqrt{s}=189-209\GeV$
with the L3 detector~\cite{l3det,lumi,alr}. This data sample comprises
$\rm 0.6\ fb^{-1}$ of integrated luminosity, as detailed in
Table~\ref{tab:1}. It extends the accessible $\sqrt{s'}$ range to
about $175\GeV$, allowing to probe Compton scattering at energies
never attained before.

%
\section{Event simulation}
%

The TEEGG Monte Carlo~\cite{teegg} is used to simulate the
$\epem\ra\epem\gamma$ process with one electron scattered in the
angular range $|\cos\theta| > 0.996$ and both the other electron and the
photon in the angular range $|\cos\theta| < 0.985$. The
$\sqrt{s'}$ spectrum for the $\rm\gamma e^{\pm} \rightarrow \gamma
e^{\pm}$ process generated by TEEGG was compared~\cite{sascha} with
the QED expectations obtained by convolving the QED cross
section with the virtual-photon flux modelled with the
equivalent-photon approach~\cite{EPA}. The two spectra agree within
the expected statistical precision of this measurement.

The following Monte Carlo programs are used to model the background
processes: BHWIDE~\cite{bhwide} for Bhabha scattering,
$\rm\epem\ra\epem$, KK2f~\cite{kk2f} for tau pair-production,
$\rm\epem\ra\tau^+\tau^-$, GGG~\cite{GGG} for multi-photon production,
$\rm\epem\ra\gamma\gamma(\gamma)$, and DIAG36~\cite{diag36} for
electron pair-production in two-photon collisions,
$\rm\epem\ra\epem\epem$.

Large samples of Monte Carlo events are generated at each $\epem$
centre-of-mass energy. The number of simulated signal events
corresponds to at least thirty times the number of expected data
events. For background processes, this factor varies from six times
for Bhabha scattering, up to more than one hundred times for the
$\rm\epem\ra\tau^+\tau^-$ process.

The L3 detector response is simulated using the GEANT
program~\cite{my_geant}, which takes into account the effects of
energy loss, multiple scattering and showering in the detector.
Time-dependent efficiencies of the different subdetectors, as
monitored during the data-taking period, are taken into account in the
simulation procedure. The simulated events are reconstructed with the
same program used for the data.

%
\section{Reconstruction of event kinematics}
%

A crucial part of the measurement is the determination of $\sqrt{s'}$
and $\cos\theta^*$. These are inferred with high precision by imposing
the constraints that there are only three particles in the final state
and one of the electrons is directed along the beam line. The
polar angles of the observed electron and photon, $\theta_{\rm
e}$ and $\theta_\gamma$, defined in Figure~\ref{fig:1}, are used to
calculate the missing energy, $E_{\rm miss}$, as:
\begin{displaymath}
	E_{\rm miss}=\sqrt{s}\frac{|\sin(\theta_{\rm e} +
	\theta_\gamma)|}{\sin\theta_{\rm e} + \sin\theta_\gamma +
	|\sin(\theta_{\rm e} + \theta_\gamma)|}.
\end{displaymath}
The square of the effective centre-of-mass energy of the $\rm \gamma e^\pm
\rightarrow \gamma e^\pm$ process is then:
\begin{displaymath}
	s'=s\bigg(1-2\frac{E_{\rm miss}}{\sqrt{s}}\bigg).
\end{displaymath}
Monte Carlo studies show that the resolution on $\sqrt{s'}$ is better
than $500\MeV$. This improves by 30\% the resolution obtained if only
the energies of the measured particles are used. The resolution on the
measured energies of the electron, $E_{\rm e}$, and of the photon
$E_{\rm \gamma}$ is also improved by re-calculating these quantities
with the kinematic constraints:
\begin{displaymath}
	E_{\rm e}^{\rm angle} =\sqrt{s}\frac{\sin\theta_\gamma}{\sin\theta_{\rm
	e} + \sin\theta_\gamma + |\sin(\theta_{\rm e} +
	\theta_\gamma)|},
\end{displaymath}
\begin{displaymath}
	E_{\gamma}^{\rm angle} =\sqrt{s}\frac{\sin\theta_{\rm e}}{\sin\theta_{\rm
	e} + \sin\theta_\gamma + |\sin(\theta_{\rm e} +
	\theta_\gamma)|}.
\end{displaymath}
Both $E_{\rm e}^{\rm angle}$ and $E_{\gamma}^{\rm angle}$ have a
resolution of about $200\MeV$. The cosine of the scattering angle of
the electron in the electron-photon centre-of-mass system is:
\begin{displaymath} 
	\cos\theta^* = \frac{\sin(\theta_{\gamma} - \theta_{\rm
	e})}{\sin\theta_{\gamma} + \sin\theta_{\rm e}}.
\end{displaymath}
The resolution on $\cos\theta^*$ in Monte Carlo events is found to be
better than 0.005.

%
\section{Event selection}
%

Events from Compton scattering must have one track in the central
tracker and two clusters in the BGO electromagnetic calorimeter in the
fiducial volume $|\cos\theta|<0.96$. The clusters must have energies
of at least $5 \GeV$ and the lateral profile of their showers must
match that expected for electromagnetic showers. One of these clusters
must be associated to the track, which should be reconstructed from at
least 80\% of the hits along its sensitive track-length. This cluster
is identified as the electron.

\begin{figure}[p]
  \begin{center}
    \begin{tabular}{cc}
      \mbox{\epsfig{figure=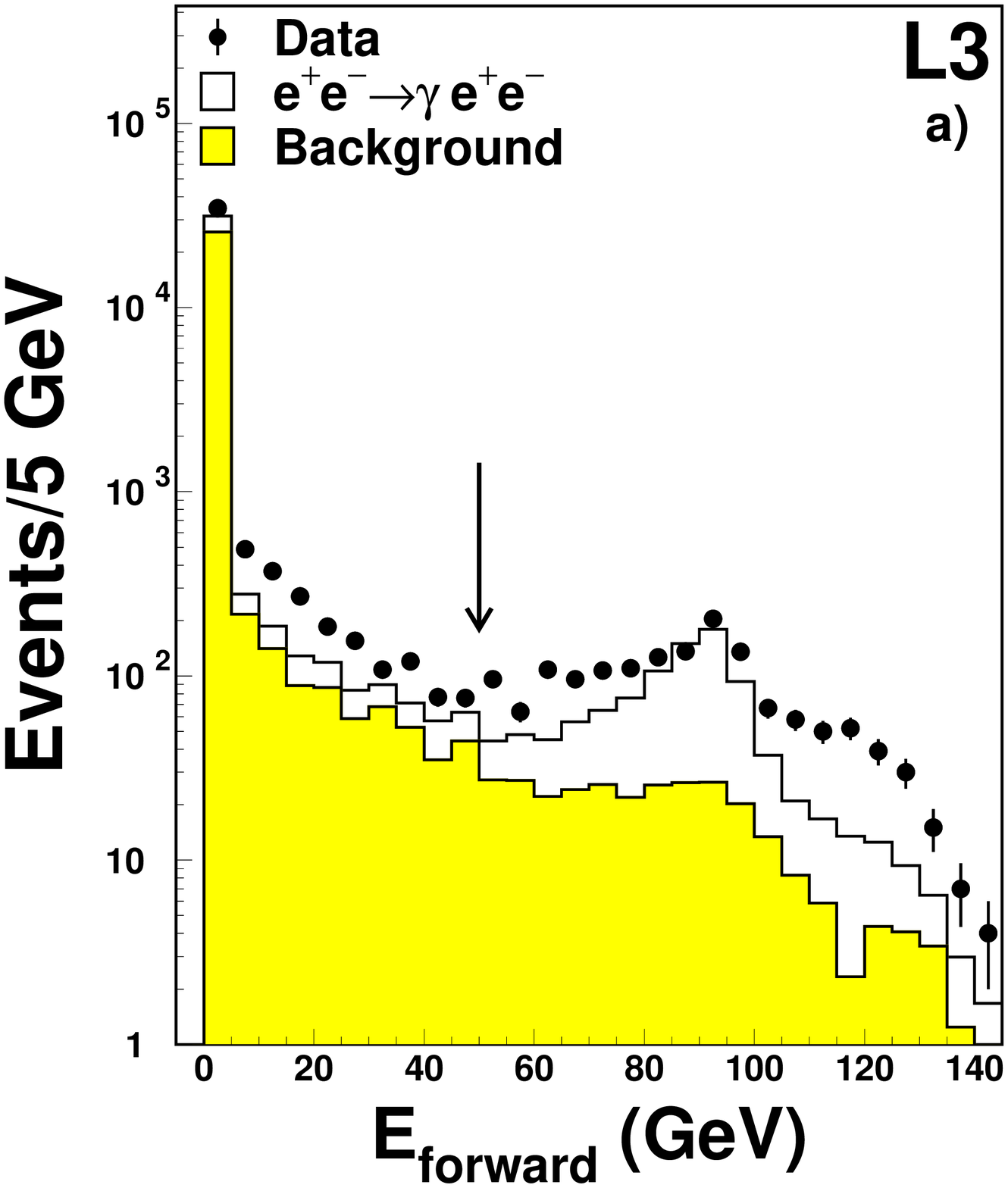,width=0.3\textwidth}}&
      \mbox{\epsfig{figure=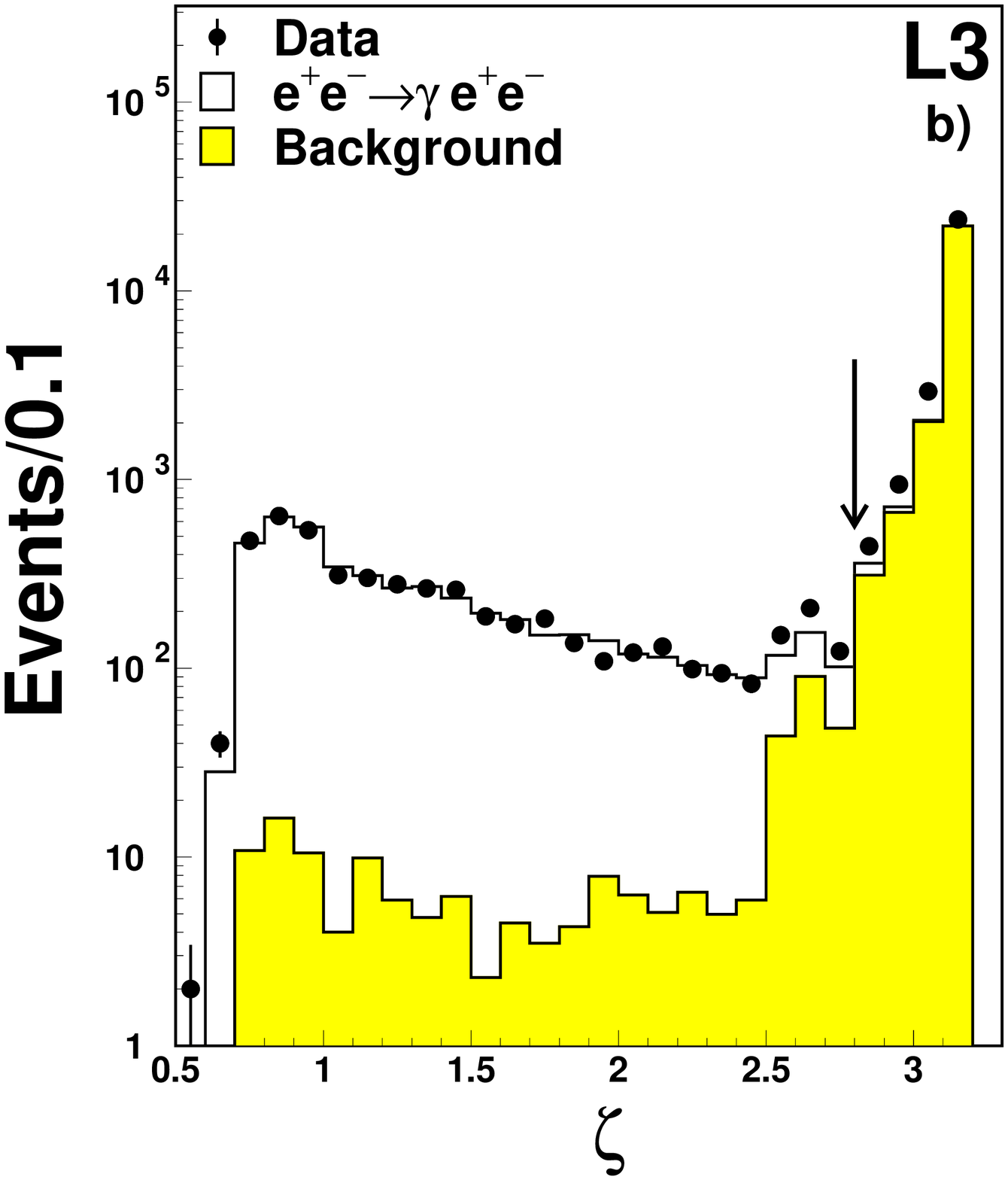,width=0.3\textwidth}}\\
      \mbox{\epsfig{figure=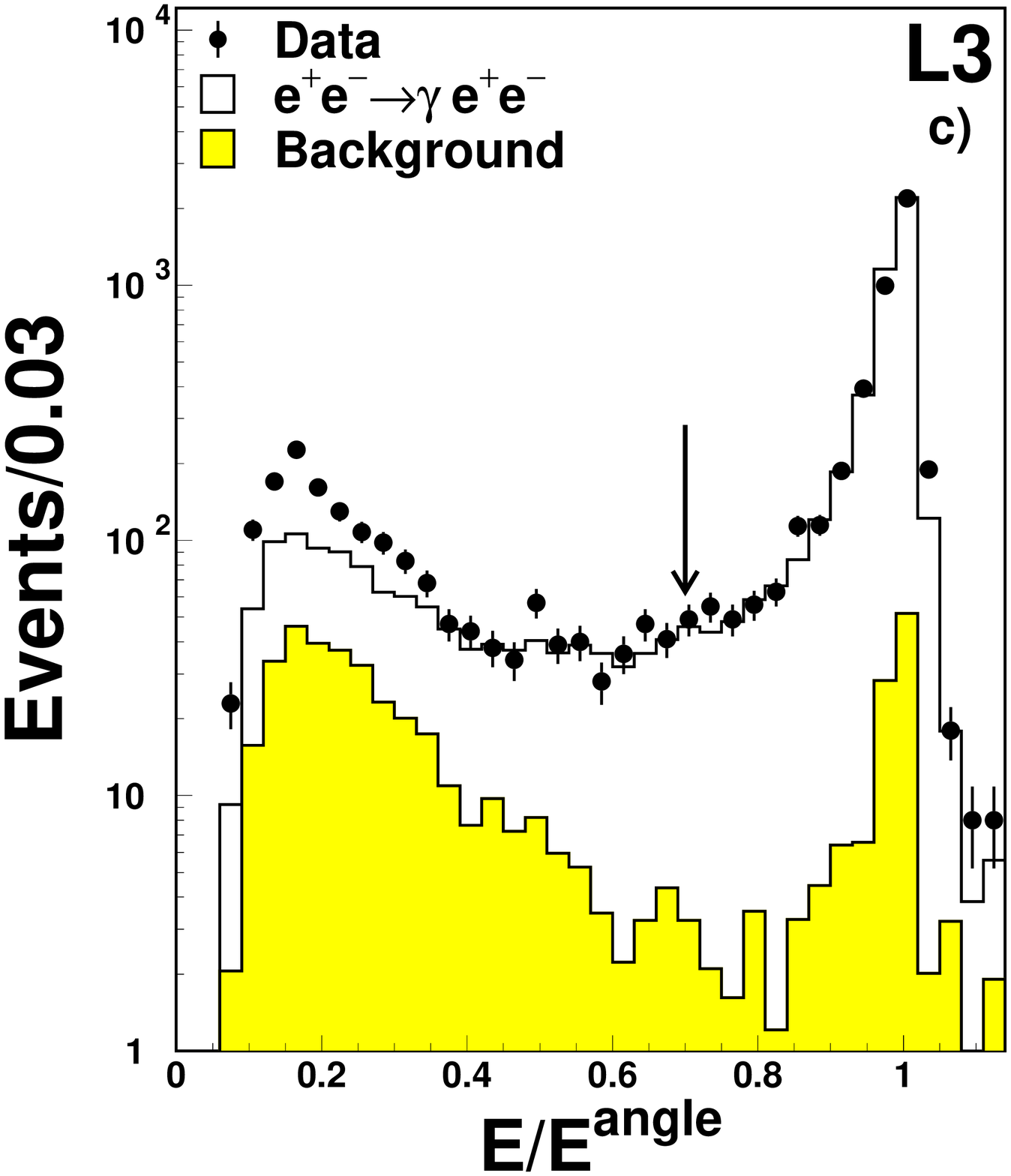,width=0.3\textwidth}}&
      \mbox{\epsfig{figure=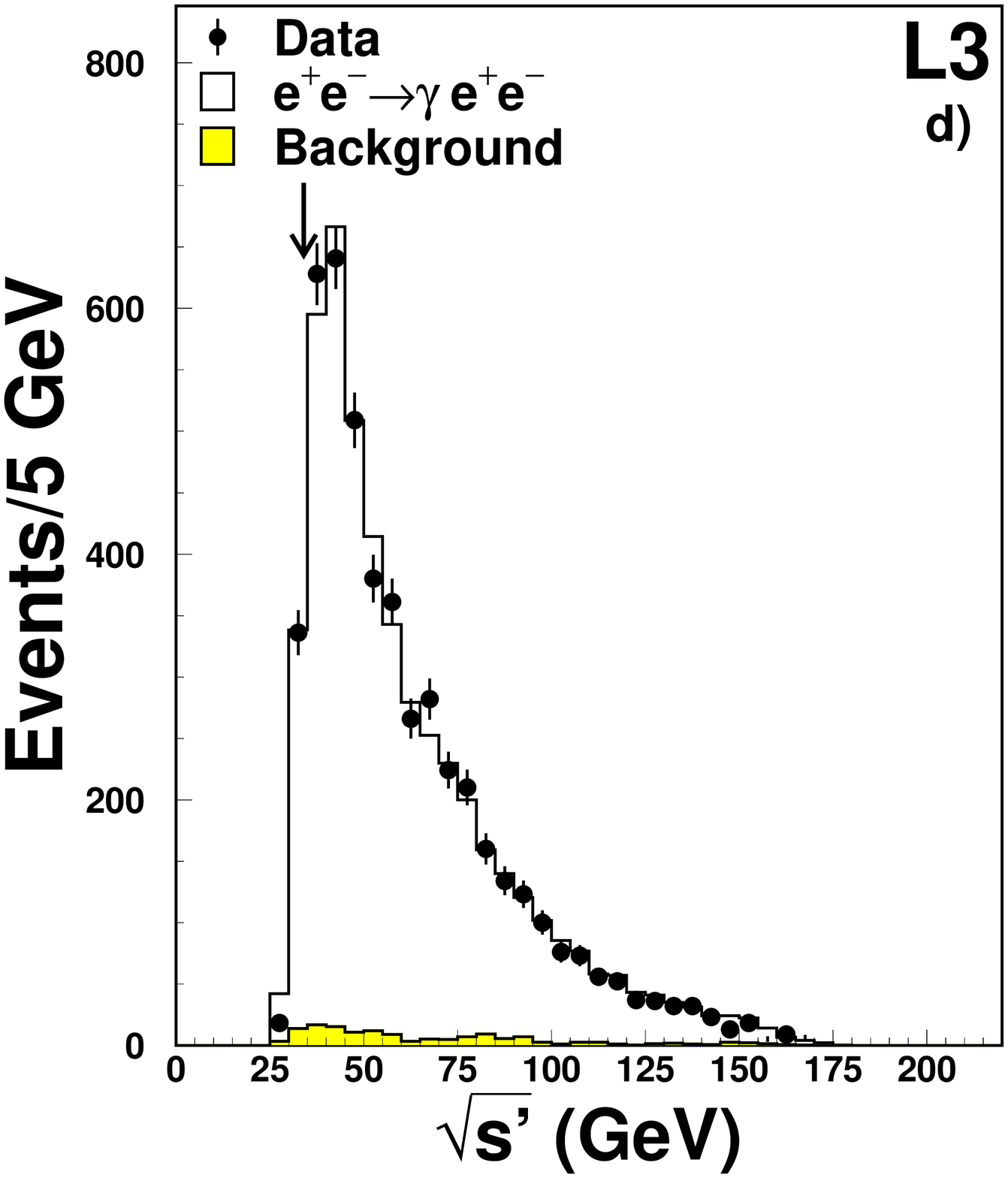,width=0.3\textwidth}}\\
      \mbox{\epsfig{figure=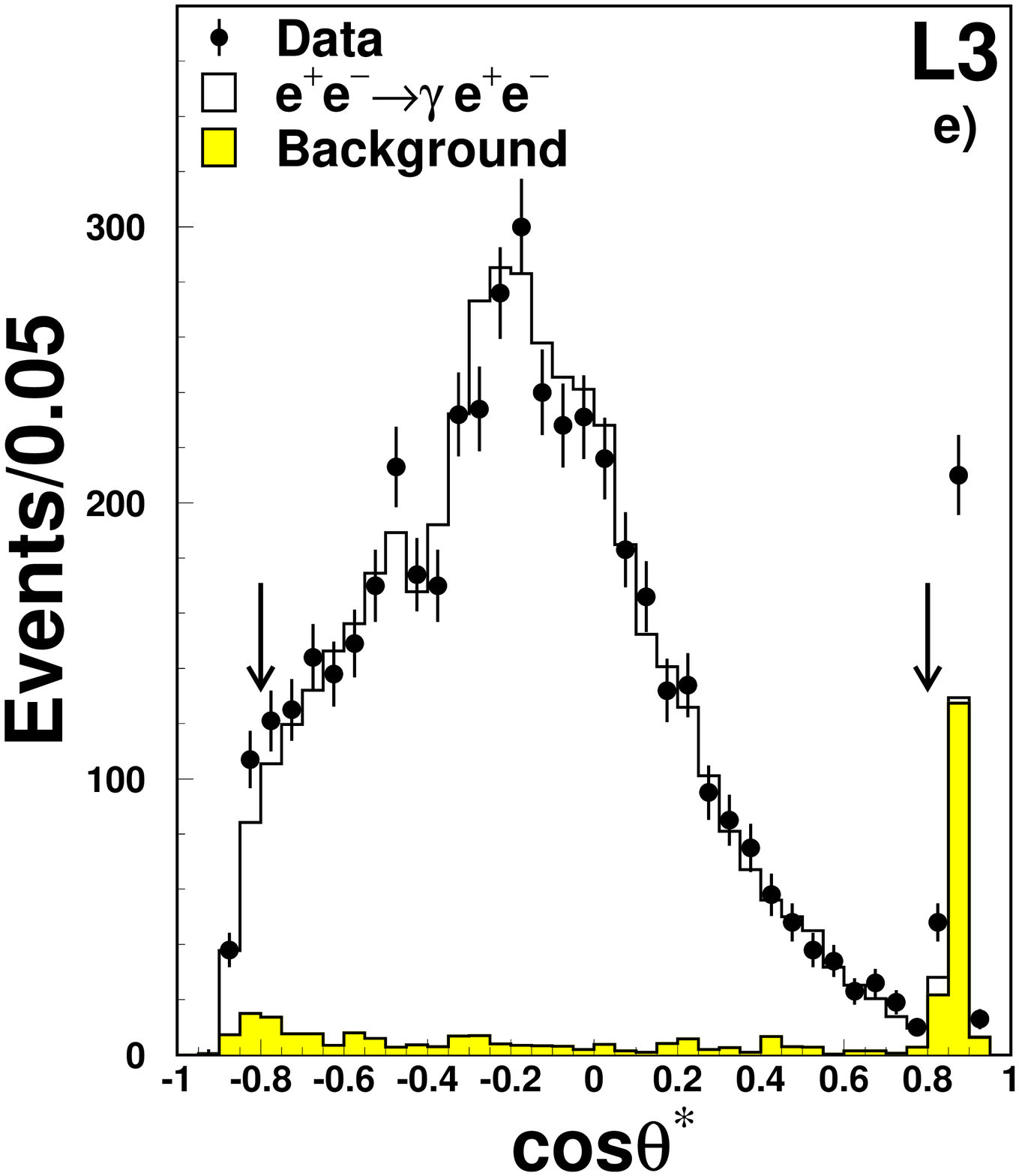,width=0.3\textwidth}}&
      \mbox{\epsfig{figure=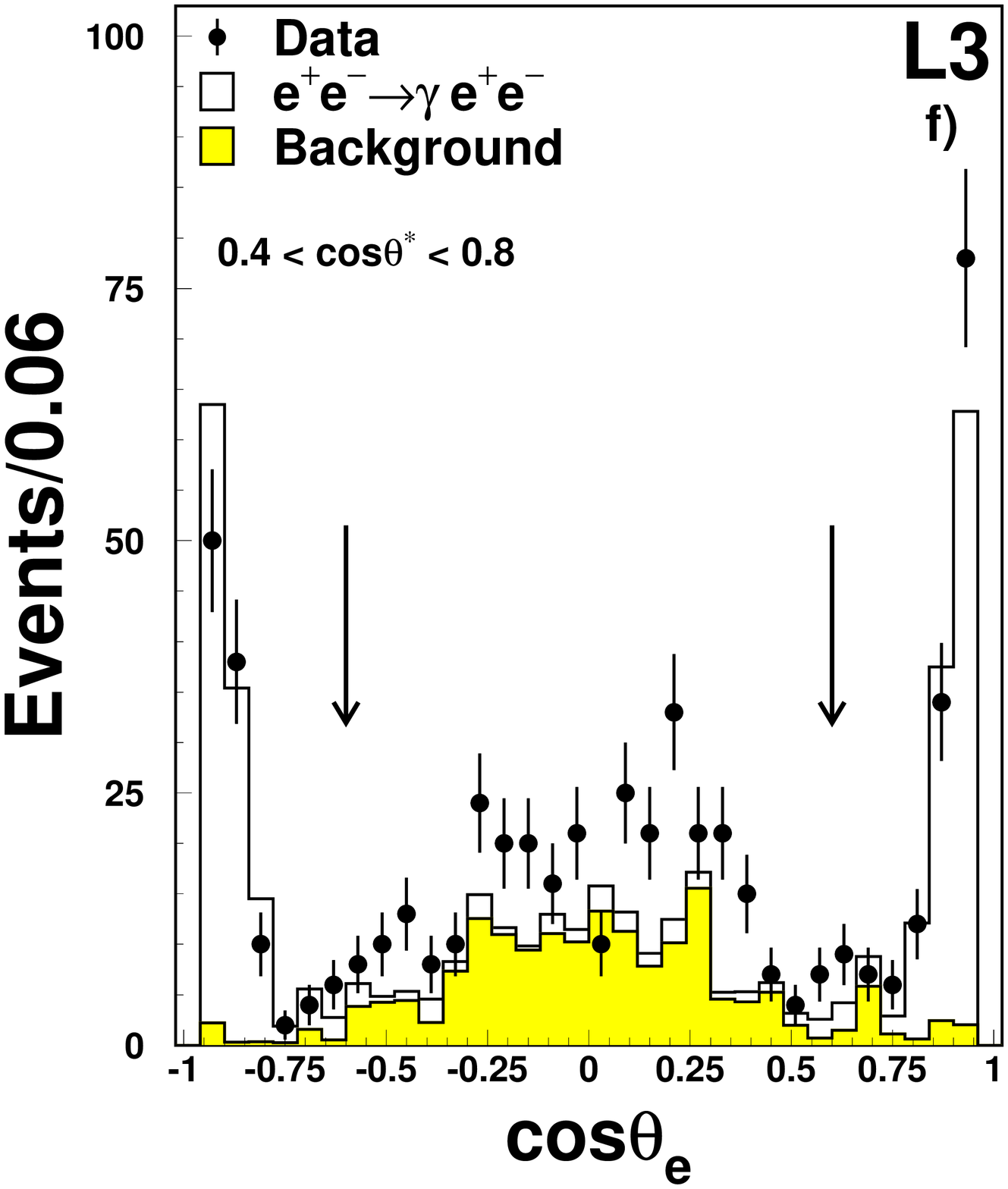,width=0.3\textwidth}}\\
    \end{tabular}
    \icaption[]{\label{fig:2} Distributions of some selection
    variables for data and Monte Carlo predictions. a) Energy in the
    forward calorimeters, $E_{\rm forward}$; b) angle between the
    electron and the photon, $\zeta$; c) ratio of the energies of the
    most energetic particle as measured in the calorimeter and as
    obtained from the angular constraints, $E/E^{\rm angle}$; d)
    effective centre-of-mass energy, $\sqrt{s'}$; e) cosine of the
    electron scattering angle in the $\rm\gamma e^\pm$ rest frame,
    $\cos\theta^*$; f) cosine of the polar angle of the electron,
    $\cos\theta_{\rm e}$, for $0.4<\cos\theta^*<0.8$. The arrows
    indicate the position of the cuts. In a) all other pre-selection
    cuts are applied. In b) all cuts are applied but the one on
    $\cos\theta^*$. In c)$-$f) cuts on all other variables are applied.}
  \end{center}
\end{figure}

The low polar-angle regions of the L3 detector are instrumented with
two calorimeters. The first is composed of BGO crystals and is used to
detect Bhabha scattering in order to measure the
luminosity~\cite{lumi}. It covers the angular region $1.4^\circ <
\theta < 3.9^\circ$. The second is built from lead and plastic
scintillators and extends this angular coverage up to
$9^\circ$~\cite{alr}. The sum of the energies deposited in these two
calorimeters, $E_{\rm forward}$, is required to be less than $50\GeV$,
as shown in Figure~\ref{fig:2}a. This cut ensures that no electron is
detected above $1.4^\circ$ and therefore only the scattering of
quasi-real photons is considered.

After these pre-selection requirements, about 36500 events are
observed in data and 32500 are expected from Monte Carlo processes, of
which 18\% is due to signal and 82\% to background. The main source of
background is Bhabha scattering with a high-energy
initial-state-radiation (ISR) photon emitted in the beam pipe, which
gives the missing-energy signature, and an electron which
mimics a photon. Electrons are misidentified as photons more frequently
in data than in Monte Carlo, which results in the excess of
observed events over the Monte Carlo prediction.  An additional
contribution to this background comes from tau pair-production where
both taus decay into electrons, one of which is identified as a
photon, and the four neutrinos are responsible for the missing-energy
signature. A lesser source of background is the production of events
with three photons, where one of the photons escapes detection along
the beam line and another, through photon conversion, is mistaken for an
electron. Several additional criteria are devised to cope with these
background sources.

Bhabha scattering and multi-photon production favour events with
electromagnetic clusters which are back-to-back in space. These
background sources are strongly reduced by a cut on the angle between the two
clusters, $\zeta<2.8$~rad, as illustrated in Figure~\ref{fig:2}b.

Events from tau pair-production are suppressed by requiring $E_{\rm
e}/E_{\rm e}^{\rm angle}> 0.7$ if $E_{\rm e}> E_{\gamma}$ or
$E_{\gamma}/E_{\gamma}^{\rm angle}> 0.7$ otherwise, as shown in
Figure~\ref{fig:2}c. This cut enforces the three-particle hypothesis
and rejects events where the missing momentum is not directed along the
beam axis.

The analysis is restricted to $\sqrt{s'}>35\GeV$, as displayed in
Figure~\ref{fig:2}d, in order to exclude the region L3 measured
with high statistics~\cite{L3Old} and concentrate on the high-energy
data.

The background from Bhabha scattering is mainly concentrated in the
forward scattering region, $\cos\theta^*>0.8$, as
presented in Figure~\ref{fig:2}e. This region is removed from the
analysis. The backward-scattering region, $\cos\theta^*<-0.8$, is also
removed in order to reduce the systematic uncertainty from a wrong
assignment of the electron charge.

The contribution from Bhabha scattering to the selected events is
further inspected. In the forward scattering region,
$0.4<\cos\theta^*<0.8$, the background electrons are mostly emitted
in the central regions of the detector, as shown in
Figure~\ref{fig:2}f. This follows from the emission of a hard ISR
photon. This background is further reduced by requiring the electron
to satisfy $|\cos\theta_{\rm e}|>0.6$ if $0.4<\cos\theta^*<0.8$.

\begin{figure}[p]
  \begin{center}
    \begin{tabular}{cc}
      \mbox{\epsfig{figure=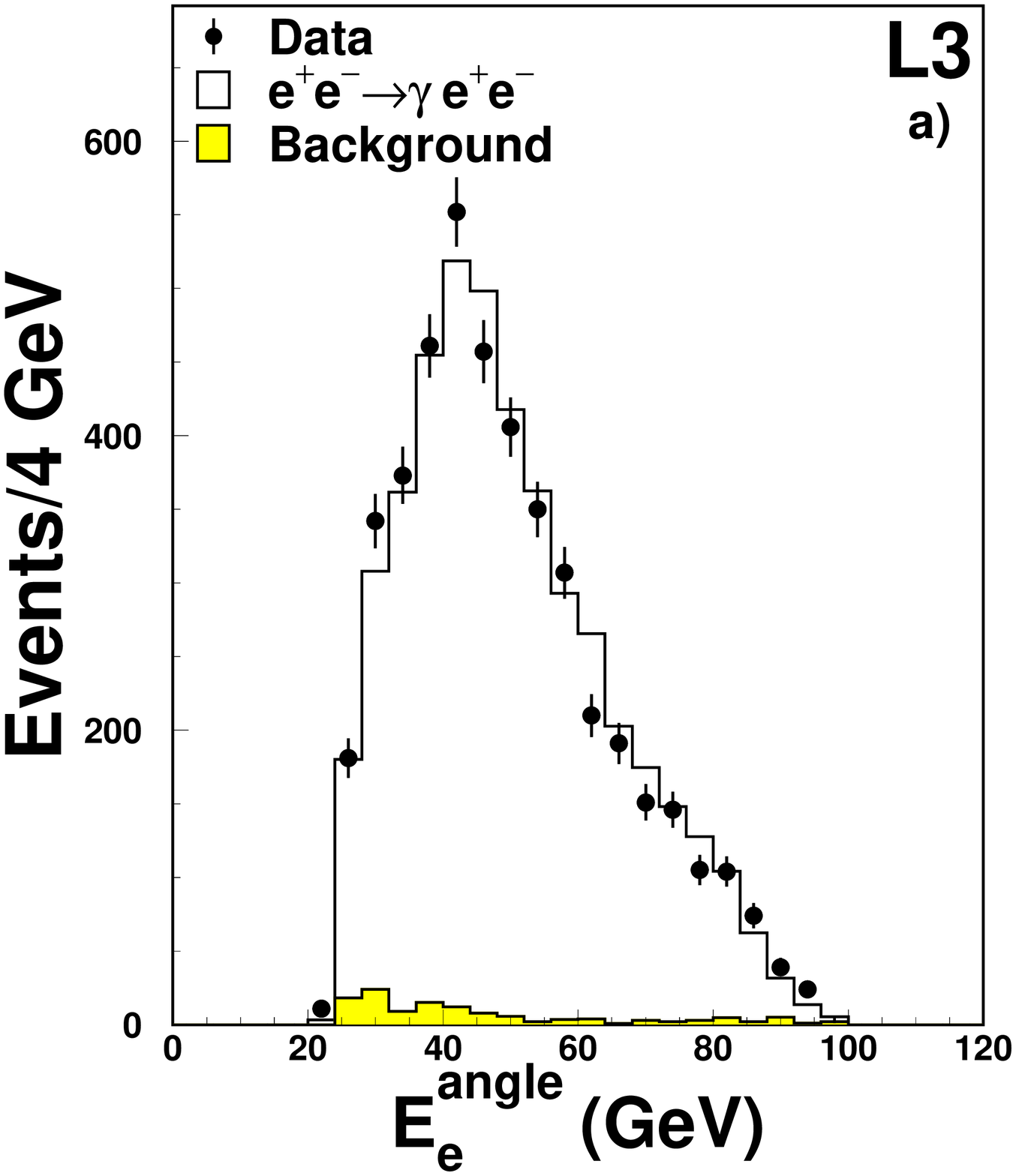,width=0.3\textwidth}}&
      \mbox{\epsfig{figure=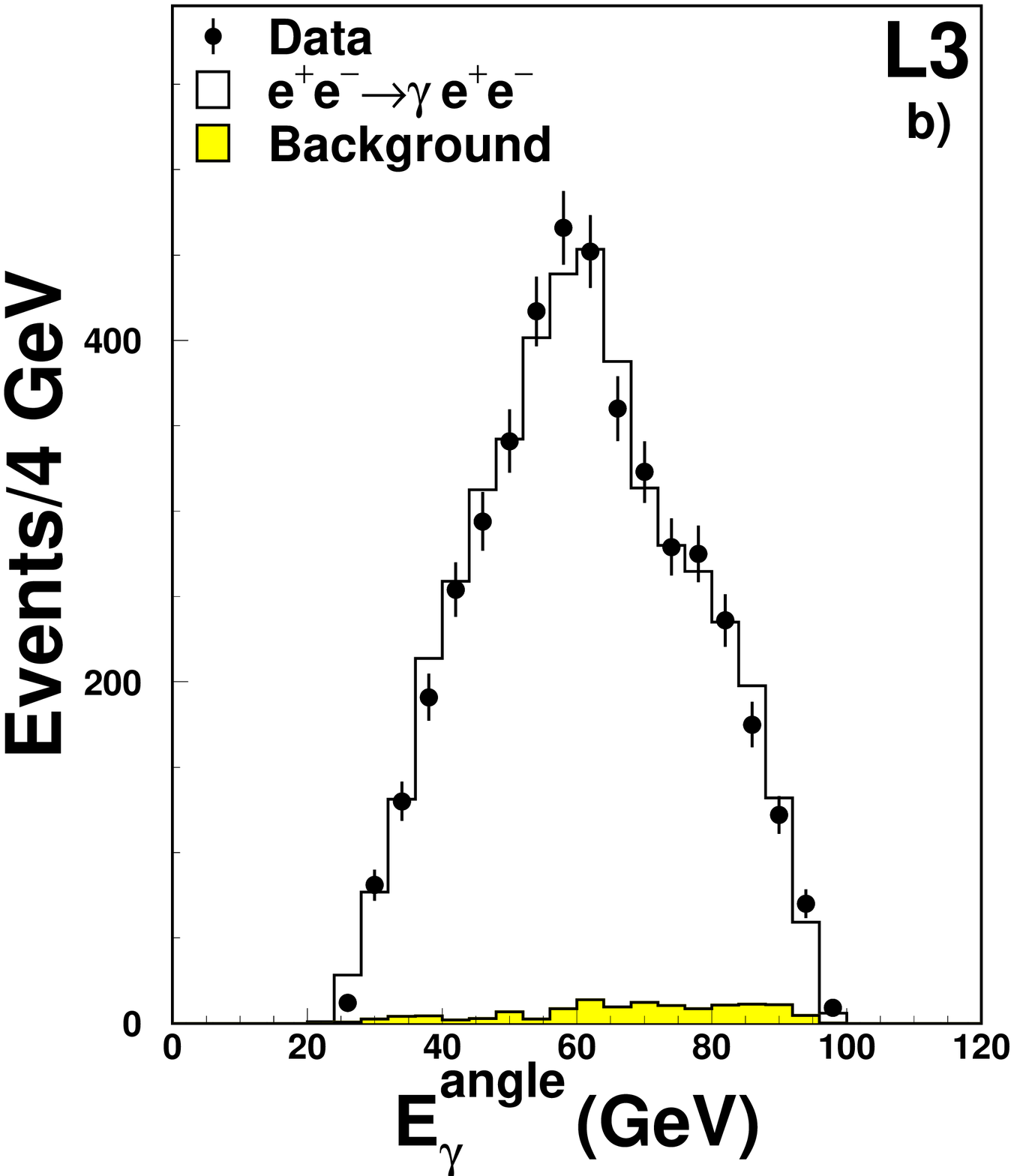,width=0.3\textwidth}}\\
      \mbox{\epsfig{figure=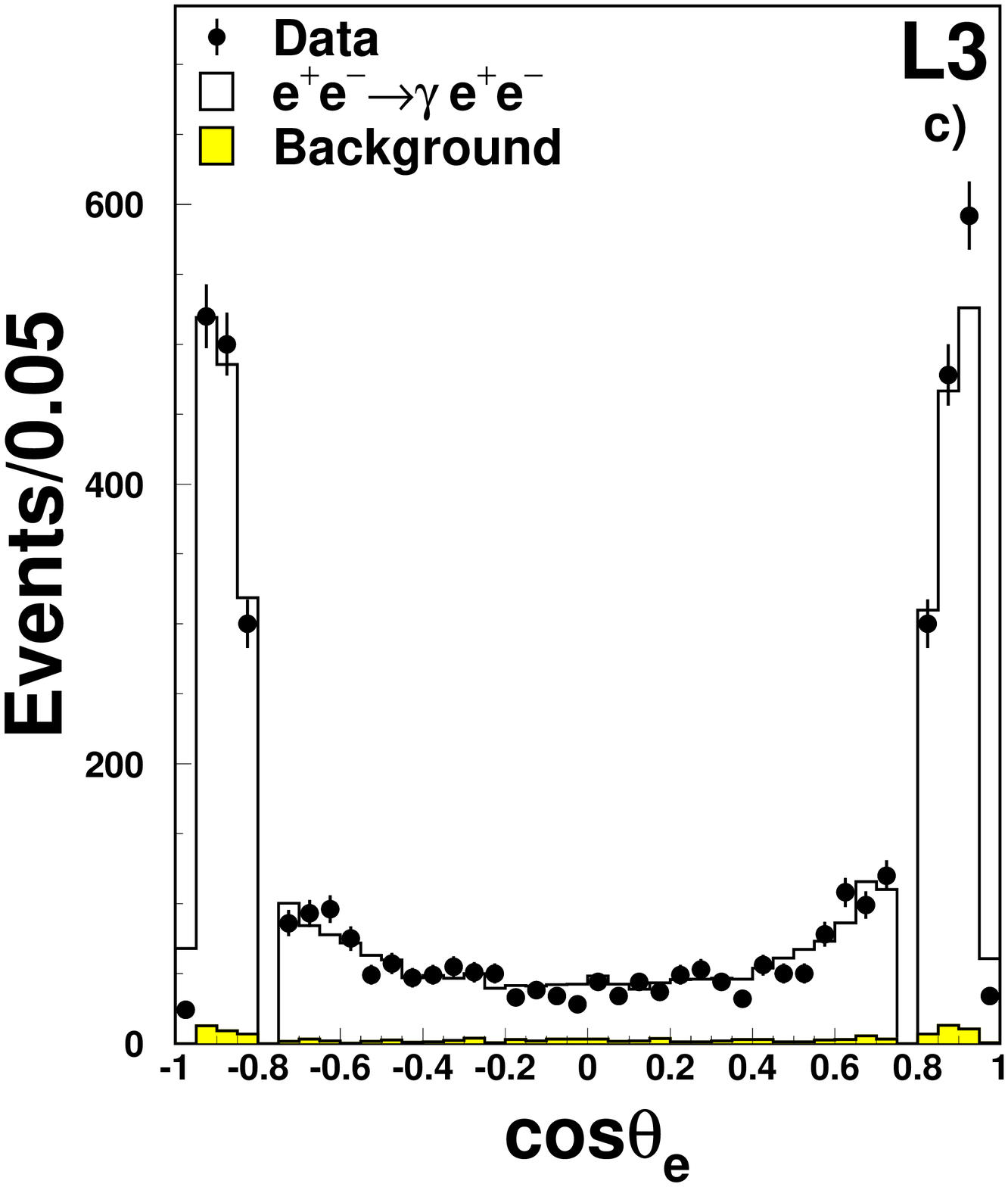,width=0.3\textwidth}}&
      \mbox{\epsfig{figure=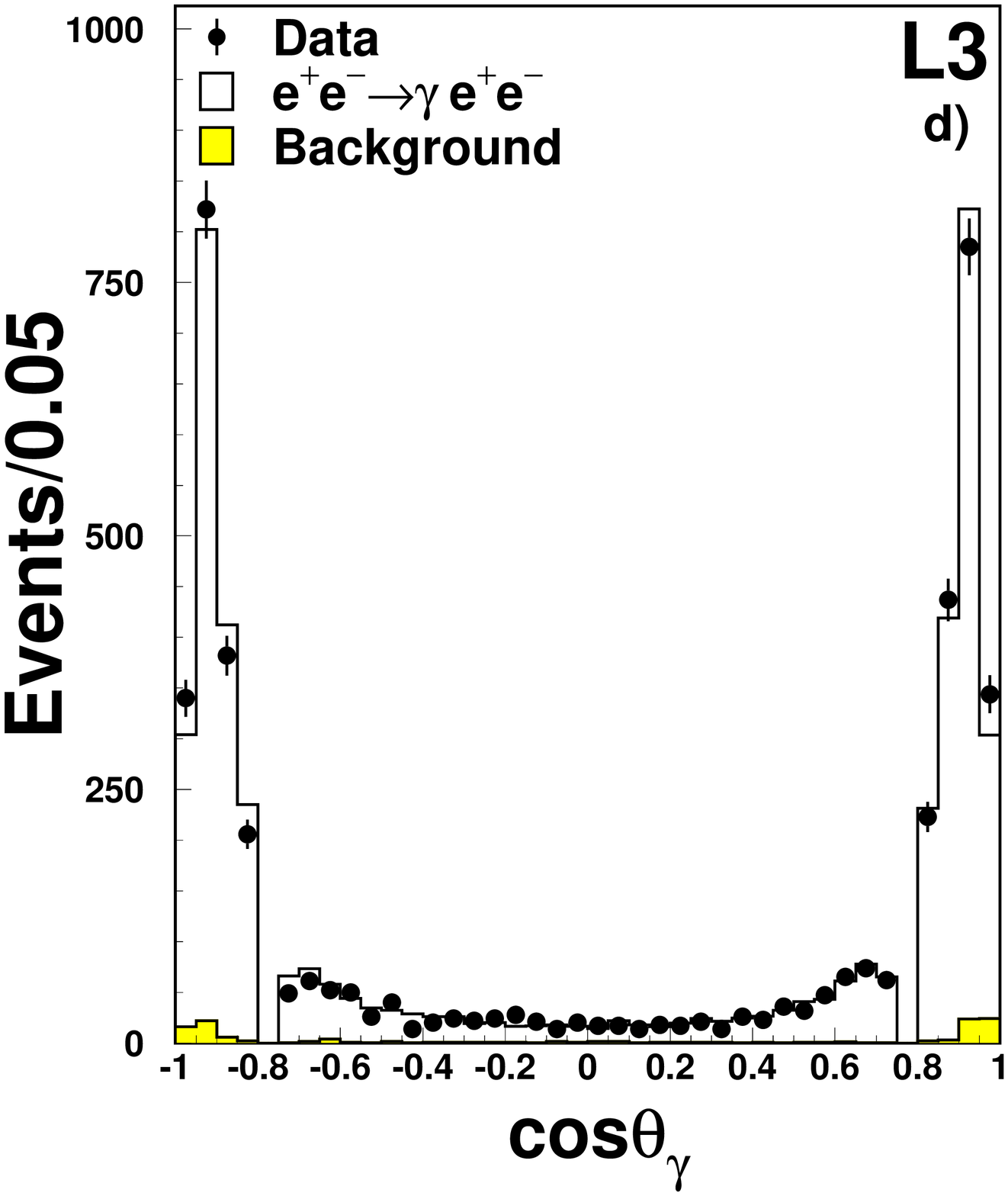,width=0.3\textwidth}}\\
      \mbox{\epsfig{figure=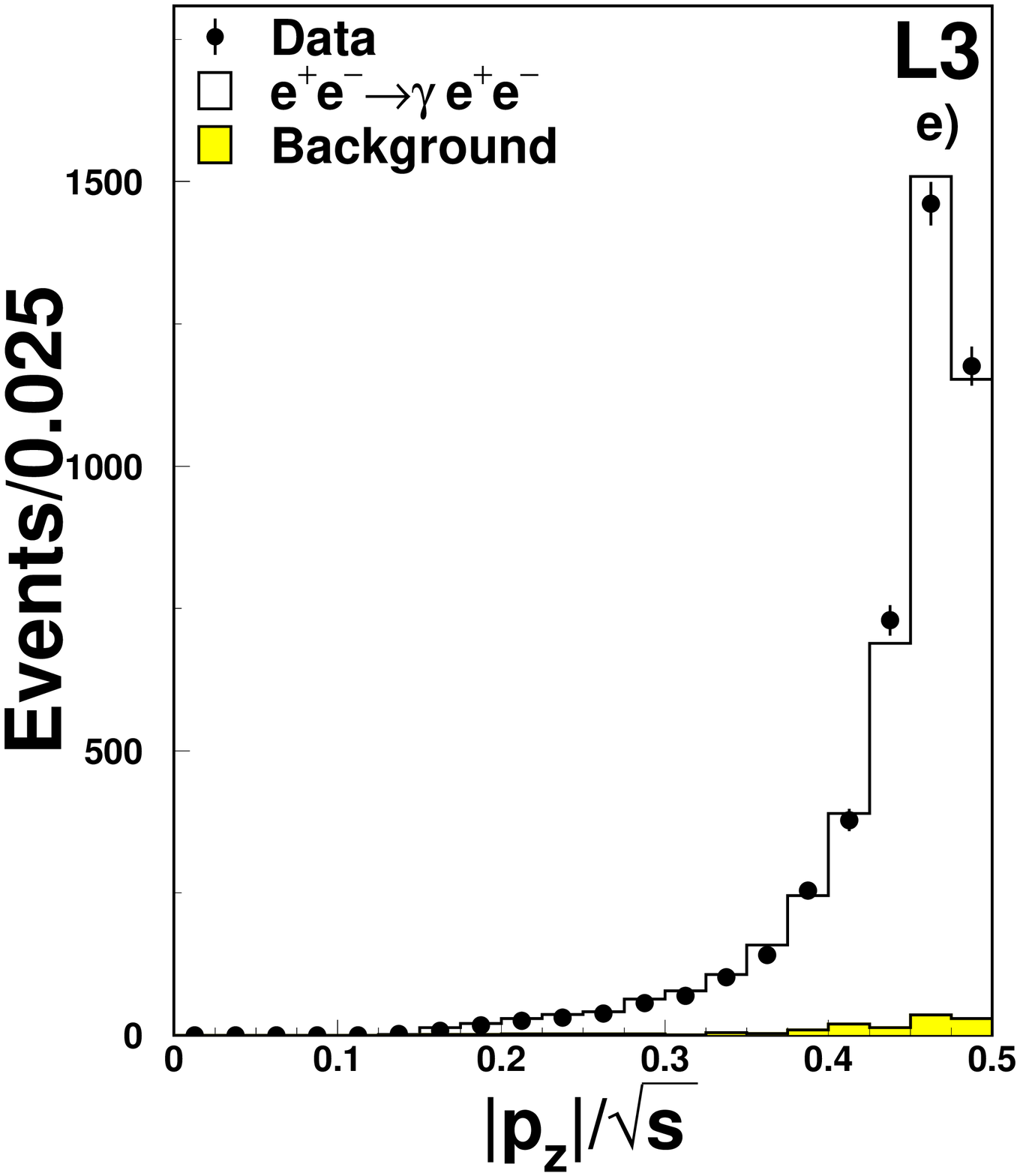,width=0.3\textwidth}}&
      \mbox{\epsfig{figure=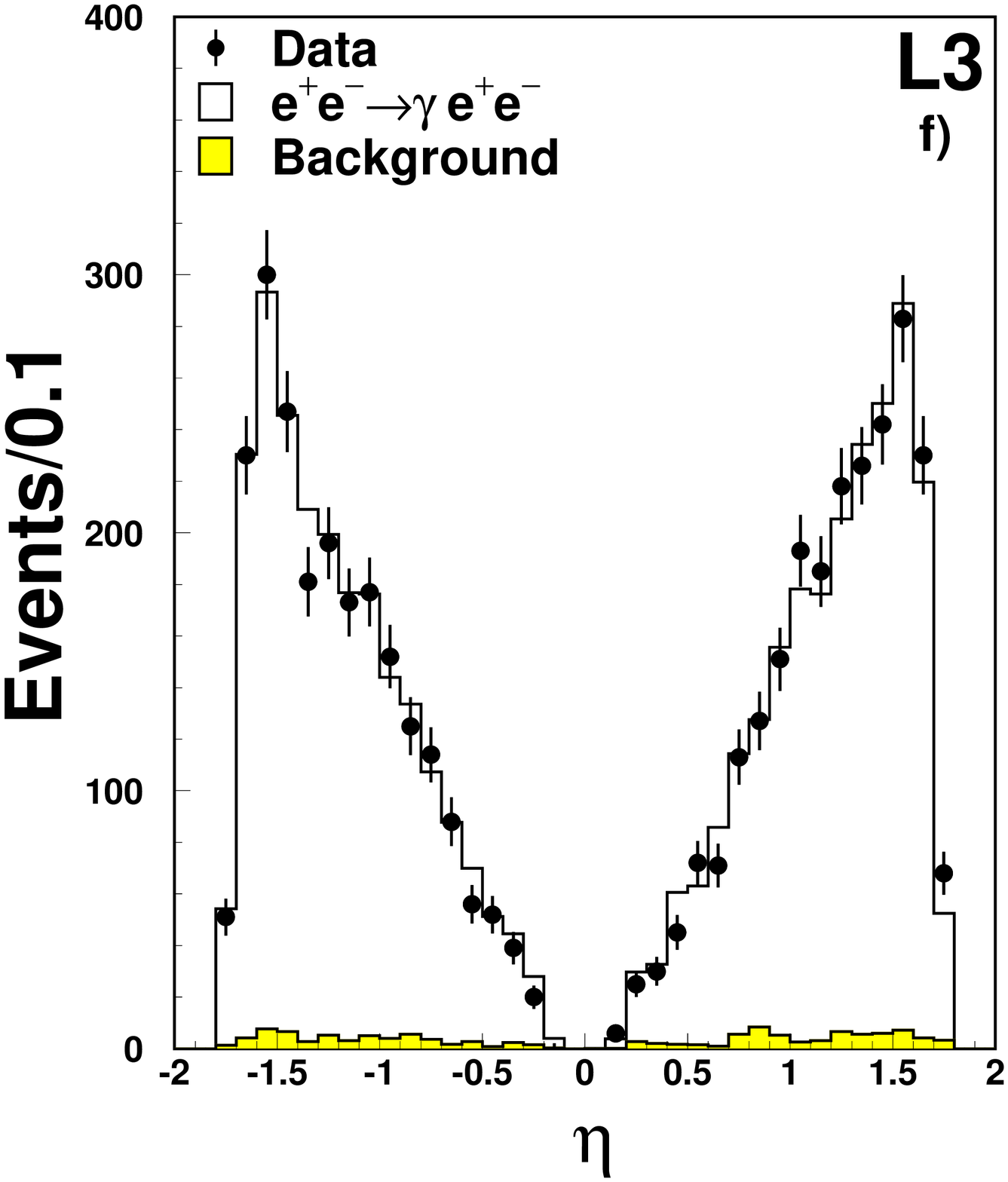,width=0.3\textwidth}}\\
    \end{tabular}
    \icaption[]{\label{fig:3} Distributions of variables for events
      selected in data and in the signal and background Monte Carlo
      samples.  Energy of a) the electron, $E_{\rm e}^{\rm angle}$,
      and b) the photon, $E_{\gamma}^{\rm angle}$; cosine of the polar
      angle of c) the electron, $\theta_{\rm e}$, and d) the photon,
      $\theta_\gamma$; e) normalised sum of the electron and photon
      longitudinal momenta, $|p_{\rm z}|/\sqrt{s}$; f) rapidity of the
      event, $\eta$.}
  \end{center}
\end{figure}

After these cuts, 4487 events are selected in data and 4534 are
expected from Monte Carlo simulations, as detailed in
Table~\ref{tab:1}~\cite{Ricardo}. Background processes are estimated
to contribute to about 3.5\% of this sample. Two thirds of the
background is due to Bhabha scattering, one quarter to electron
production in two photon collisions and the rest to multi-photon
production and tau pair-production.  The distributions of the energies
and the angles of the electrons and photons of the selected events are
shown in Figures~\ref{fig:3}a--d. Figure~\ref{fig:3}e presents the
normalised sum of the momenta of the electron and photon along the
beam line, $|p_{\rm z}|/\sqrt{s}$. As expected, a large boost along
the beam line is observed. Figure~\ref{fig:3}f shows the rapidity of
the selected events.

Monte Carlo studies indicate that the average value of $Q^2$ for the
selected events is $0.48 \GeV^2$ and 90\% of the photons satisfy
$Q^2<2\GeV^2$. This corresponds to a small average-virtuality $\langle
Q^2/s' \rangle \simeq 3\times 10^{-3}$, which justifies the treatment
of the virtual photons as quasi-real ones.  The kinematics of the
process under investigation is such that $p_{\rm t}^2\simeq Q^2$,
where $p_{\rm t}$ is the sum of the momenta of the final state
electron and photon in the plane transverse to the
beams. Figure~\ref{fig:4}a shows the distribution of $p_{\rm t}^2$ for
the selected events. An average value $<p_{\rm t}^2>=0.3\GeV^2$ is
observed, with a root-mean-square of $1\GeV^2$, which further
validates the hypothesis of quasi-real photons.

The distribution of $\cos\theta^*$ for the selected events, shown in
Figure~\ref{fig:4}b, presents the characteristic
backward-scattering behaviour of Compton
scattering. Figure~\ref{fig:4}c displays the observed and expected
distributions of $\sqrt{s'}$. The average value of $\sqrt{s'}$ is
$64\GeV$.  Ten events are observed in data for $\sqrt{s'} > 163\GeV$,
up to $\sqrt{s'} =175\GeV$, the largest energies at which Compton
scattering was ever observed. Monte Carlo simulations predict $9\pm1$
events in this region, with a background of 7\%. The uncertainty is
due to the limited statistics of the generated Monte Carlo sample.
Figure~\ref{fig:4} presents one of these high-energy events.

%
\section{Systematic uncertainties}
%

Several sources of systematic uncertainty are considered and their
impact on the measurement of the cross section as a function of
$\sqrt{s'}$ and of the differential cross section as a function of
$\cos\theta^*$ are listed in Table~\ref{tab:2}.

The measurement of the photon and electron angles is crucial to the
determination of the event kinematics. These variables are smeared by
$\pm1\%$ to account for possible uncertainties in detector alignment,
time-dependent changes in resolution or discrepancies between the data
and the Monte Carlo simulations. The effects of these changes are
larger for events with large values of $\sqrt{s'}$ or $\cos\theta^*$.
In addition, the energy scale of the electromagnetic calorimeter is
varied within its uncertainty of $\pm1\%$. As the energies are mainly
inferred from the angular measurements, this change has a small impact
on the cross sections.

The amount of charge confusion in the tracker is measured with control
data-samples and is well reproduced in the Monte Carlo
simulations. However, uncertainties in this simulation are a potential
source of systematics. This is assessed by injecting in the
simulations an additional amount of charge confusion. A small
variation in the cross section as a function of $\sqrt{s'}$ is
observed, whereas a larger uncertainty affects the determination of
the differential cross section for large values of $\cos\theta^*$.

Uncertainties of the background normalisation are also propagated to
the final results. All background levels are varied by $\pm2\%$ with
the exception of electron production in two-photon collisions, varied
by $\pm10\%$.

Finally, the effects of the limited amount of signal and background
Monte Carlo statistics are treated as additional systematic
uncertainties.

%
\section{Results}
%

\begin{figure}[p]
  \begin{center}
    \begin{tabular}{c}
      \mbox{\epsfig{figure=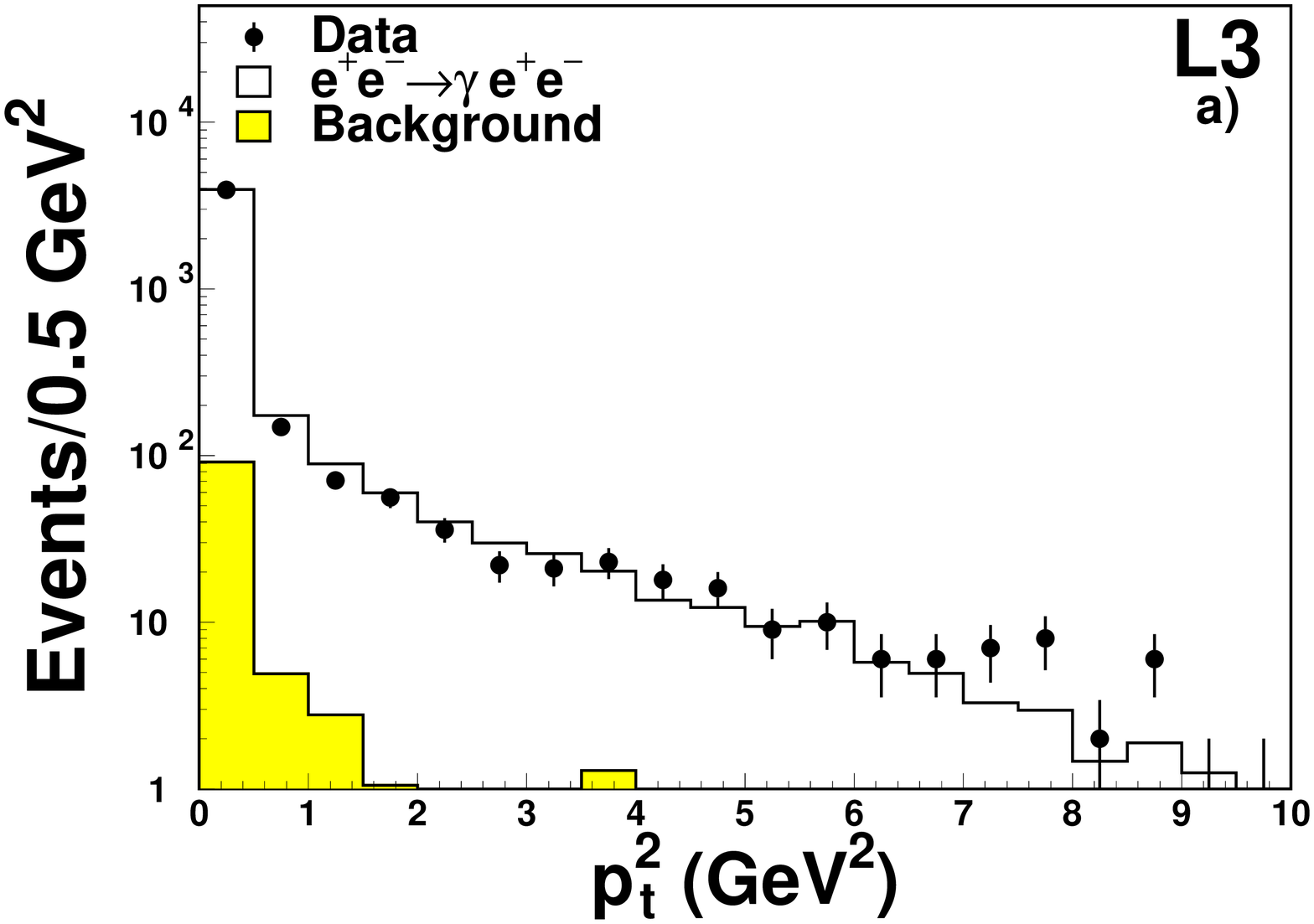,width=0.5\textwidth}}\\
      \mbox{\epsfig{figure=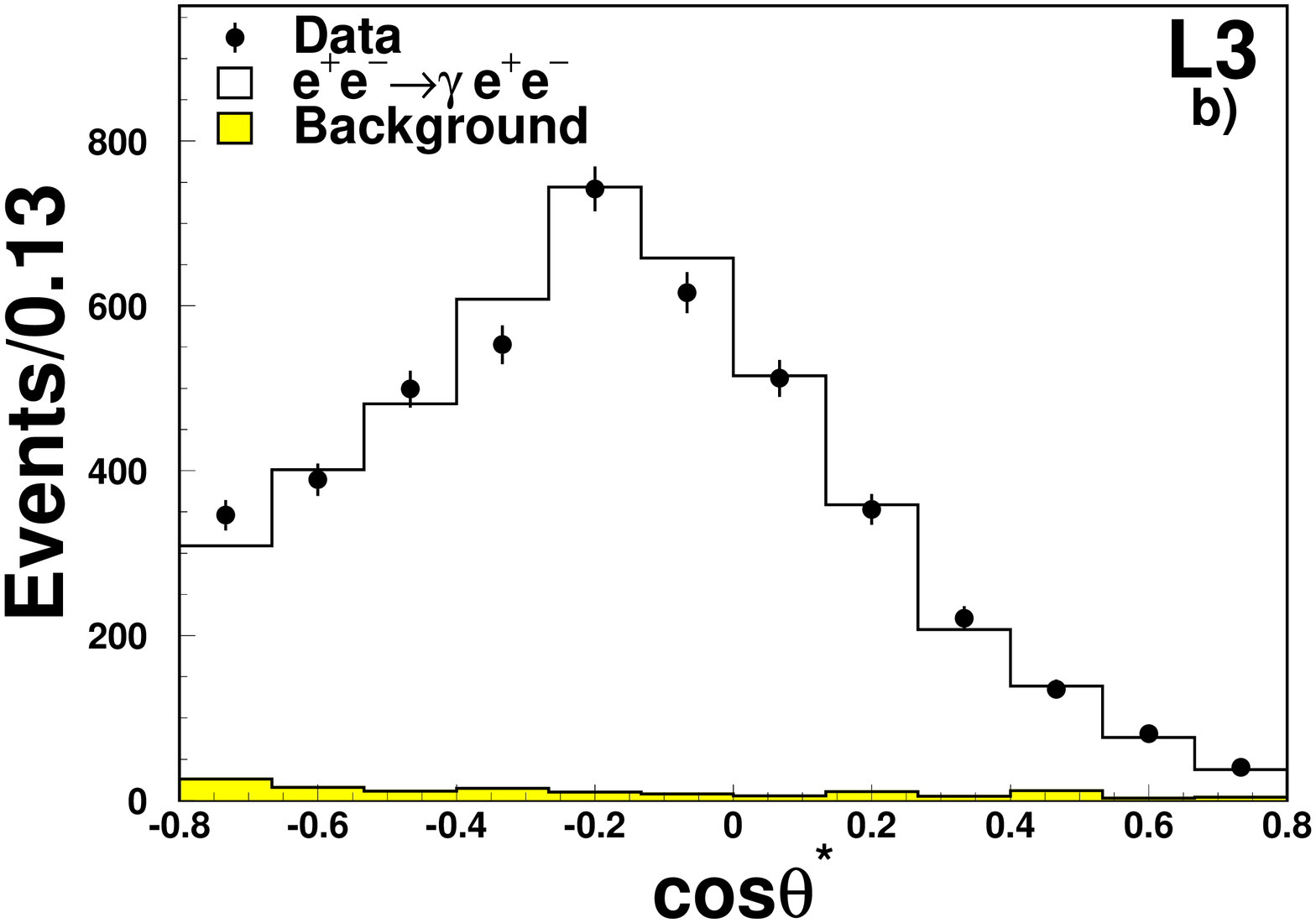,width=0.5\textwidth}}\\
      \mbox{\epsfig{figure=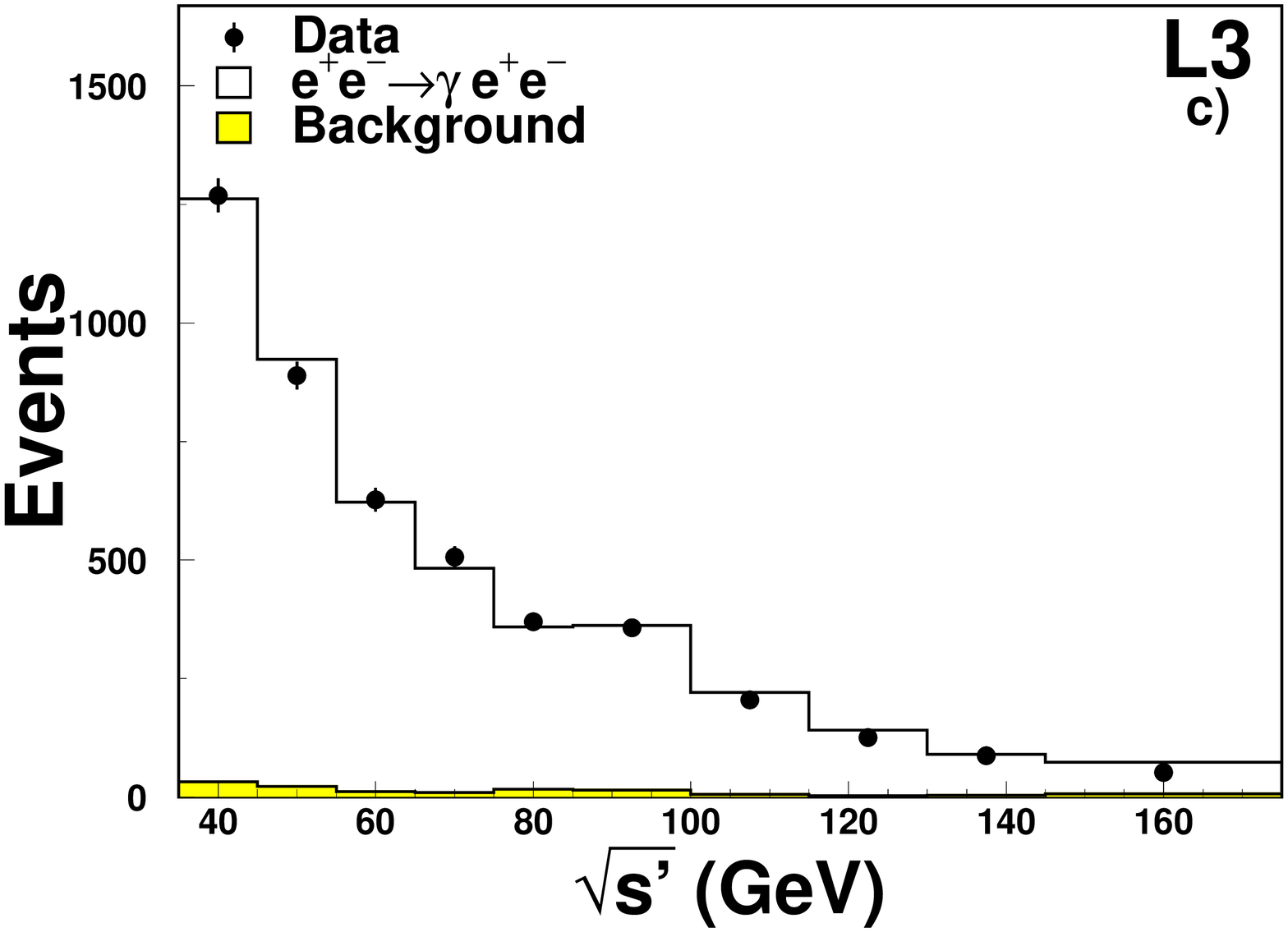,width=0.5\textwidth}}\\
    \end{tabular}
    \icaption[]{\label{fig:4} Distributions of variables for events
      selected in data and in the signal and background Monte Carlo
      samples. a) Square of the sum of the electron and photon
      transverse momenta, $p_{\rm t}^2$; b) cosine of the electron
      scattering angle in the $\rm\gamma e^\pm$ rest frame,
      $\cos\theta^*$; c) effective centre-of-mass energy,
      $\sqrt{s'}$.}
  \end{center}
\end{figure}

The phase space covered by this analysis, $35\GeV<\sqrt{s'}<175\GeV$
and $-0.8<\cos\theta^*<0.8$, is divided in ten intervals in
$\sqrt{s'}$ and twelve intervals in $\cos\theta^*$. The limits of
these intervals are detailed in Tables~\ref{tab:3} and~\ref{tab:4},
together with their average values\footnote{The average values are
calculated as suggested in Reference~\citen{Average}.}  and the
numbers of observed and expected events, also shown in
Figures~\ref{fig:4}b~and~\ref{fig:4}c.

The differential cross section of the $\epem\ra\epem\gamma$ process as
a function of $\sqrt{s'}$ is related to the cross section of the $\rm
\gamma e^\pm\ra \gamma e^\pm$ process by:
\begin{equation}
	{{\rm d}\sigma_{\epem\ra\epem\gamma} \over {\rm d}\sqrt{s'} }
	= f_\gamma(\sqrt{s},\sqrt{s'}) \, \sigma_{\rm \gamma e^\pm\ra
	\gamma e^\pm} ( \sqrt{s'} ),
\label{eq:2}
\end{equation}
where $f_\gamma(\sqrt{s},\sqrt{s'})$ is the virtual-photon flux. The
cross section of the $\epem\ra\epem\gamma$ process for the $i$-th
interval of $\sqrt{s'}$ can be extracted from the numbers of events
observed in data and expected from signal and background Monte Carlo
simulations, $N_{\rm data}(i)$, $N^{\rm sign}_{\rm MC}(i)$ and $N^{\rm
back}_{\rm MC}(i)$, respectively, as:
\begin{equation}
 {{\Delta}\sigma_{\rm \epem\ra\epem\gamma} \over {\Delta}\sqrt{s'} } =
  {{\Delta} \sigma^{\rm MC}_{\epem\ra\epem\gamma}\over
  {\Delta}\sqrt{s'} } \, \frac{N_{\rm data}(i) - N^{\rm back}_{\rm
  MC}(i)}{N^{\rm sign}_{\rm MC}(i)},
\label{eq:3}
\end{equation}
where ${{\Delta}\sigma_{\epem\ra\epem\gamma}^{\rm MC}
/{\Delta}\sqrt{s'} }$ is the cross section predicted by the Monte
Carlo.  By expressing Equation~\ref{eq:3} in terms of
Equation~\ref{eq:2}, and assuming that $f_\gamma(\sqrt{s},\sqrt{s'})$
is the same for data and Monte Carlo, the cross section for quasi-real
Compton scattering at the average effective centre-of-mass energy
$\langle \sqrt{s'} \rangle$ can be derived directly from the values in
Table~\ref{tab:3} as:
\begin{equation}
 \sigma_{\rm \gamma e^\pm\ra \gamma e^\pm} \left( \langle \sqrt{s'}
  \rangle \right) = \sigma^{\rm QED}_{\rm \gamma e^\pm\ra\gamma e^\pm
  }\left( \langle \sqrt{s'} \rangle \right) \frac{N_{\rm data}(i) -
  N^{\rm back}_{\rm MC}(i)}{N^{\rm sign}_{\rm MC}(i)},
  \label{eq:4}
\end{equation}
where $\sigma^{\rm QED}_{\rm \gamma e^\pm \ra \gamma e^\pm}\left(
\langle \sqrt{s'} \rangle \right)$ is the value expected from QED. The
differential cross section for quasi-real Compton scattering as a
function of $\cos\theta^*$ is derived from a formula equivalent to
Equation~\ref{eq:4}, {\it mutatis mutandis}.

\begin{figure}[p]
  \begin{center}
    \begin{tabular}{cc}
      \mbox{\epsfig{figure=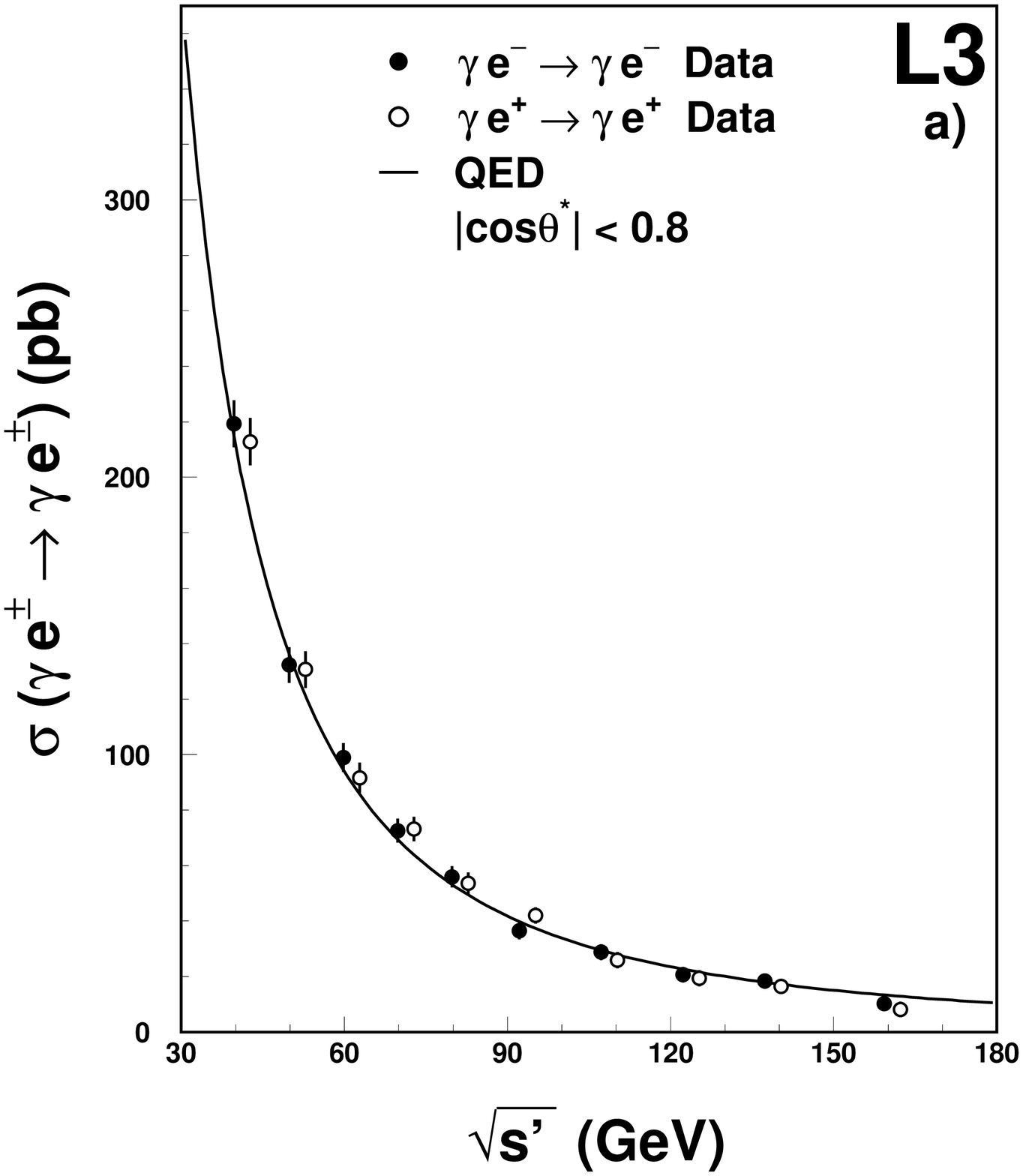,width=0.4\textwidth}} &
      \mbox{\epsfig{figure=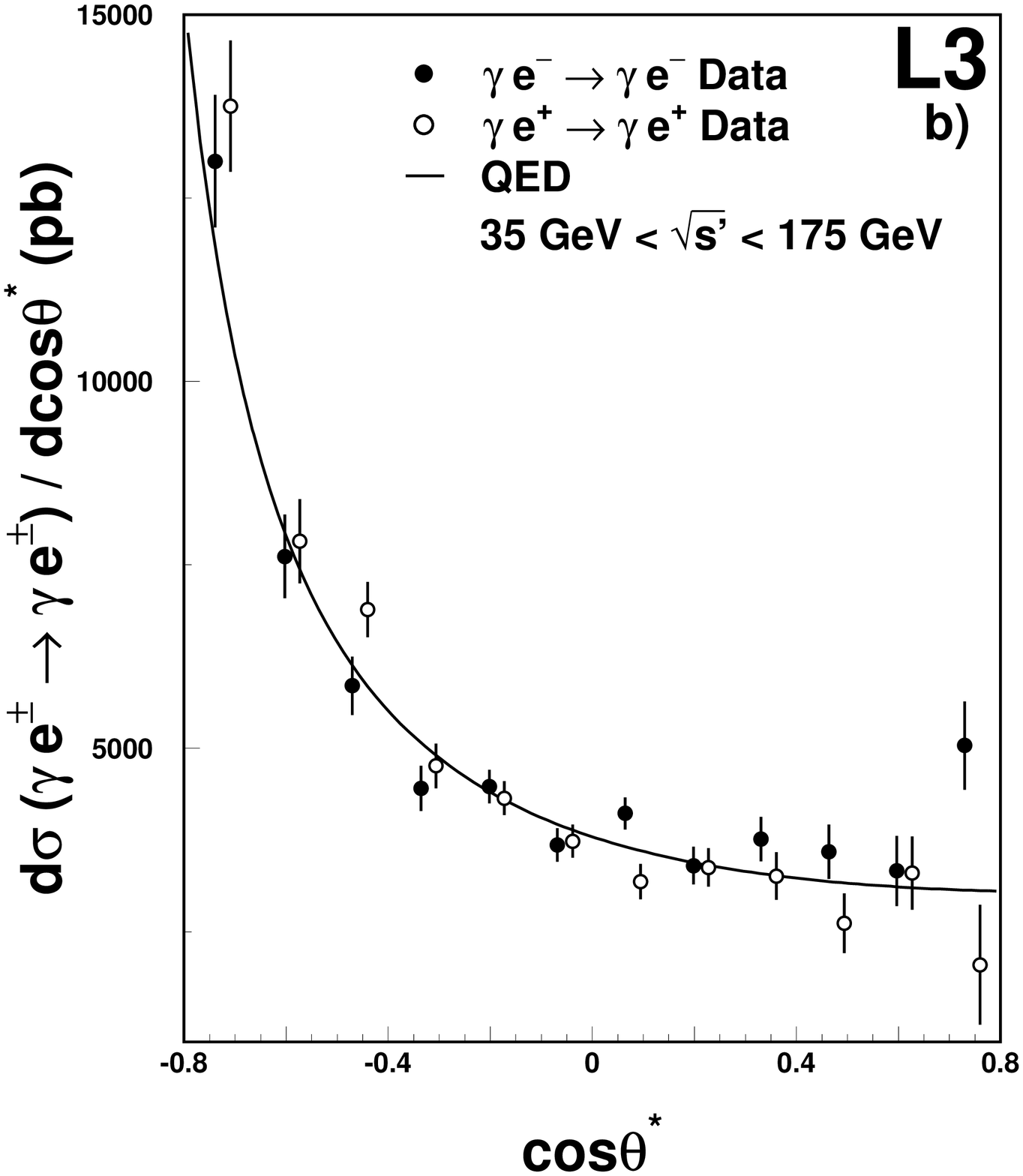,width=0.4\textwidth}} \\
      \mbox{\epsfig{figure=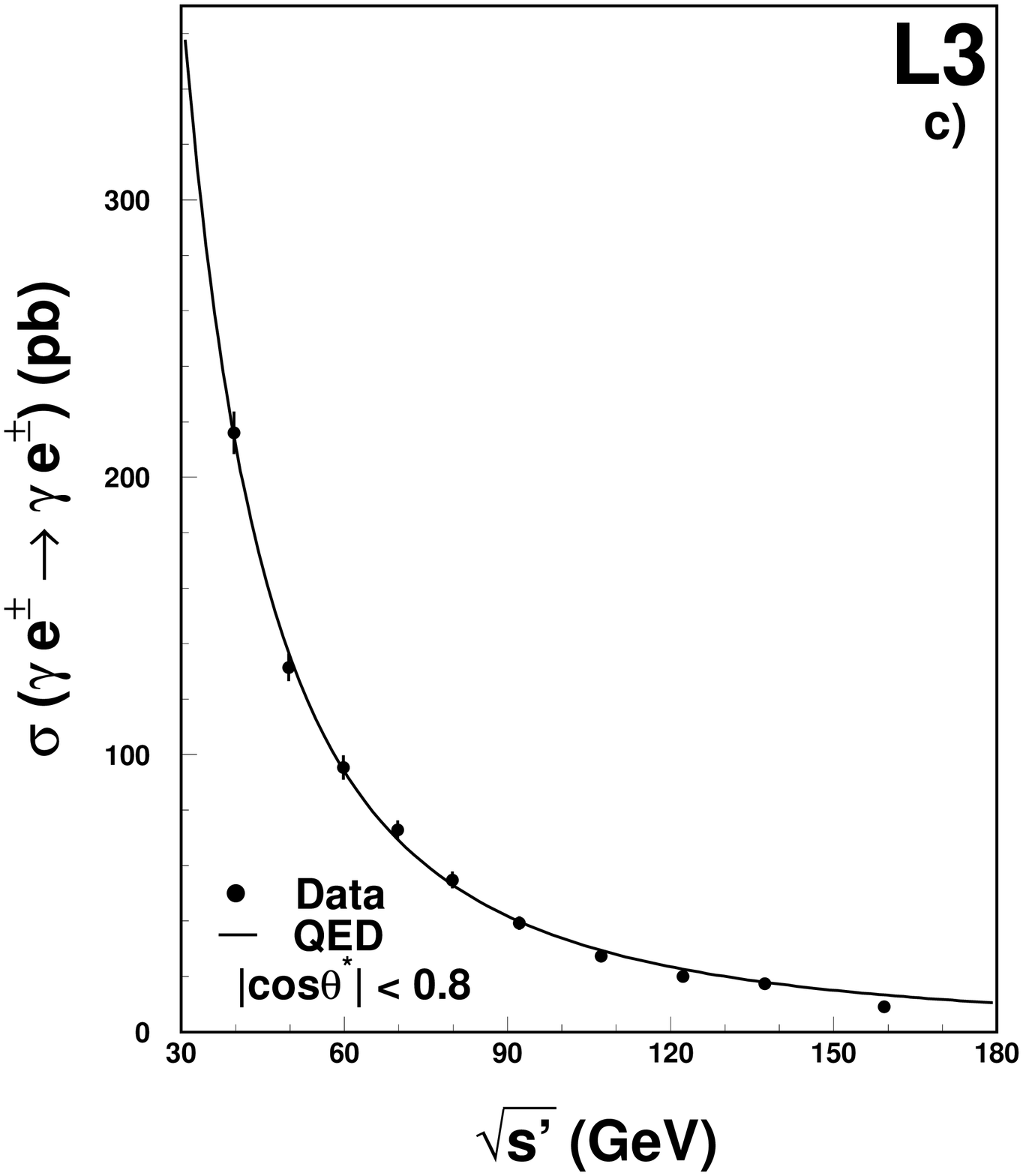,width=0.4\textwidth}} &
      \mbox{\epsfig{figure=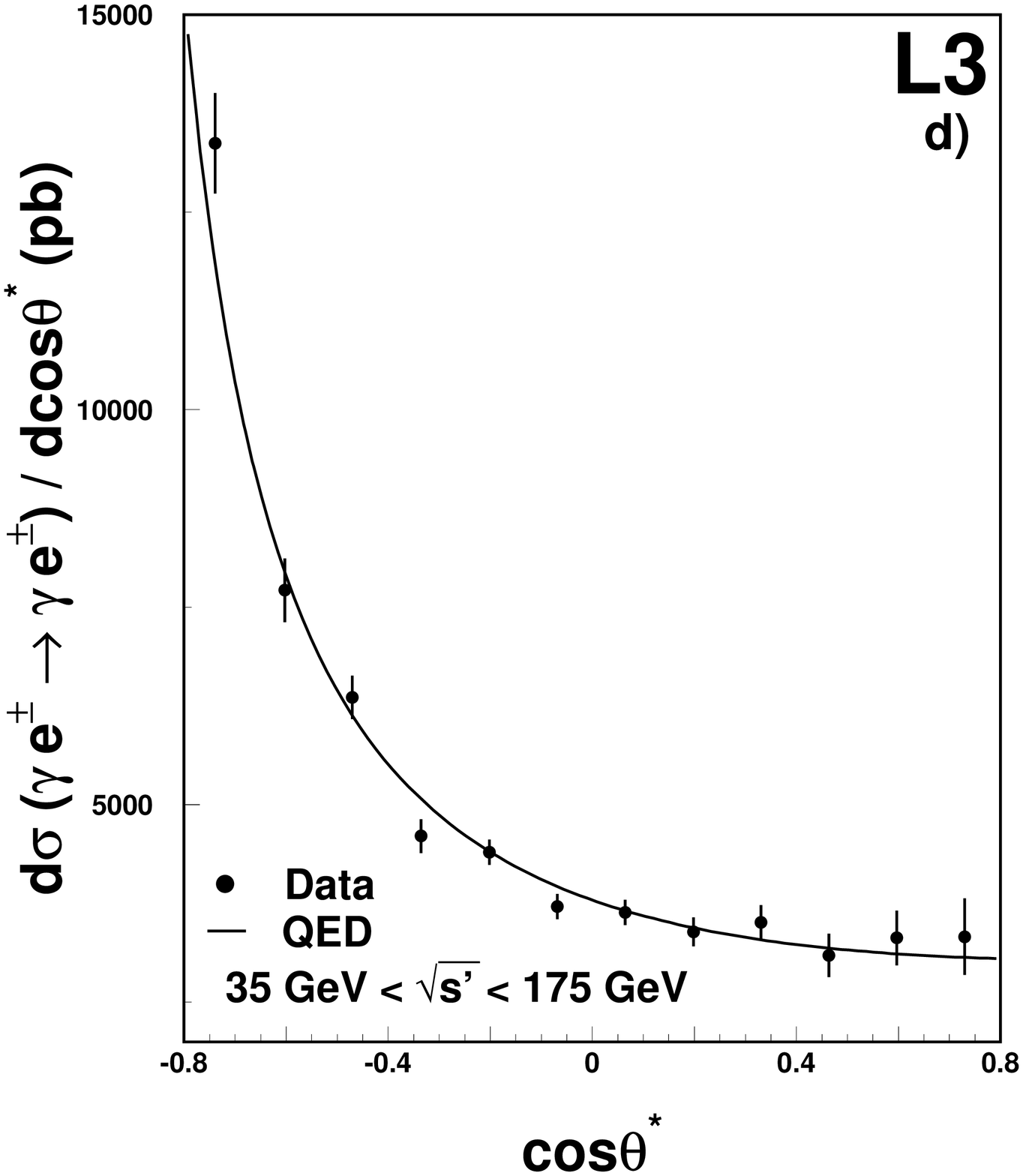,width=0.4\textwidth}} \\
    \end{tabular}
    \icaption[]{\label{fig:6}
      Measured cross sections of the $\rm\gamma e^+\ra\gamma e^+$,
      $\rm\gamma e^-\ra\gamma e^-$ and $\rm\gamma e^{\pm}\ra\gamma e^{\pm}$
      process as a function of a) and c) the effective centre-of-mass
      energy for $|\cos\theta^*|<0.8$ and b) and d) the rest-frame
      scattering angle for $\sqrt{s'}=35-175\GeV$. For clarity, the
      measurements for positrons in a) and b) are slightly displaced
      to the
      right.  Data collected at $\epem$ centre-of-mass energies
      $\sqrt{s}=188.6-209.2\GeV$ are considered and the QED predictions are
      also shown.
    }
  \end{center}
\end{figure}

The selected events are further classified according to the presence
of either electrons or positrons in the final states. The cross
section as a function of $ \sqrt{s'}$ and the differential cross
section as a function of $\cos\theta$ are measured and the results are
presented in Figures~\ref{fig:6}a~and~\ref{fig:6}b. They are in good
mutual agreement.  The combined results for electrons and positrons
are presented in Tables~\ref{tab:3} and~\ref{tab:4} and in
Figures~\ref{fig:6}c~and~\ref{fig:6}d. All results are in good
agreement with the QED predictions, also presented in
Tables~\ref{tab:3} and~\ref{tab:4} and in Figure~\ref{fig:6}. The
predictions for the cross section as a function of $ \sqrt{s'}$ are
derived by integrating Equation~\ref{eq:1} over the range
$-0.8<\cos\theta^*<0.8$, while the predictions for the differential
cross section as a function of $\cos\theta$ are derived by integrating
over the range $35\GeV<\sqrt{s'}<175\GeV$.

\begin{figure}[p]
  \begin{center}
    \mbox{\epsfig{figure=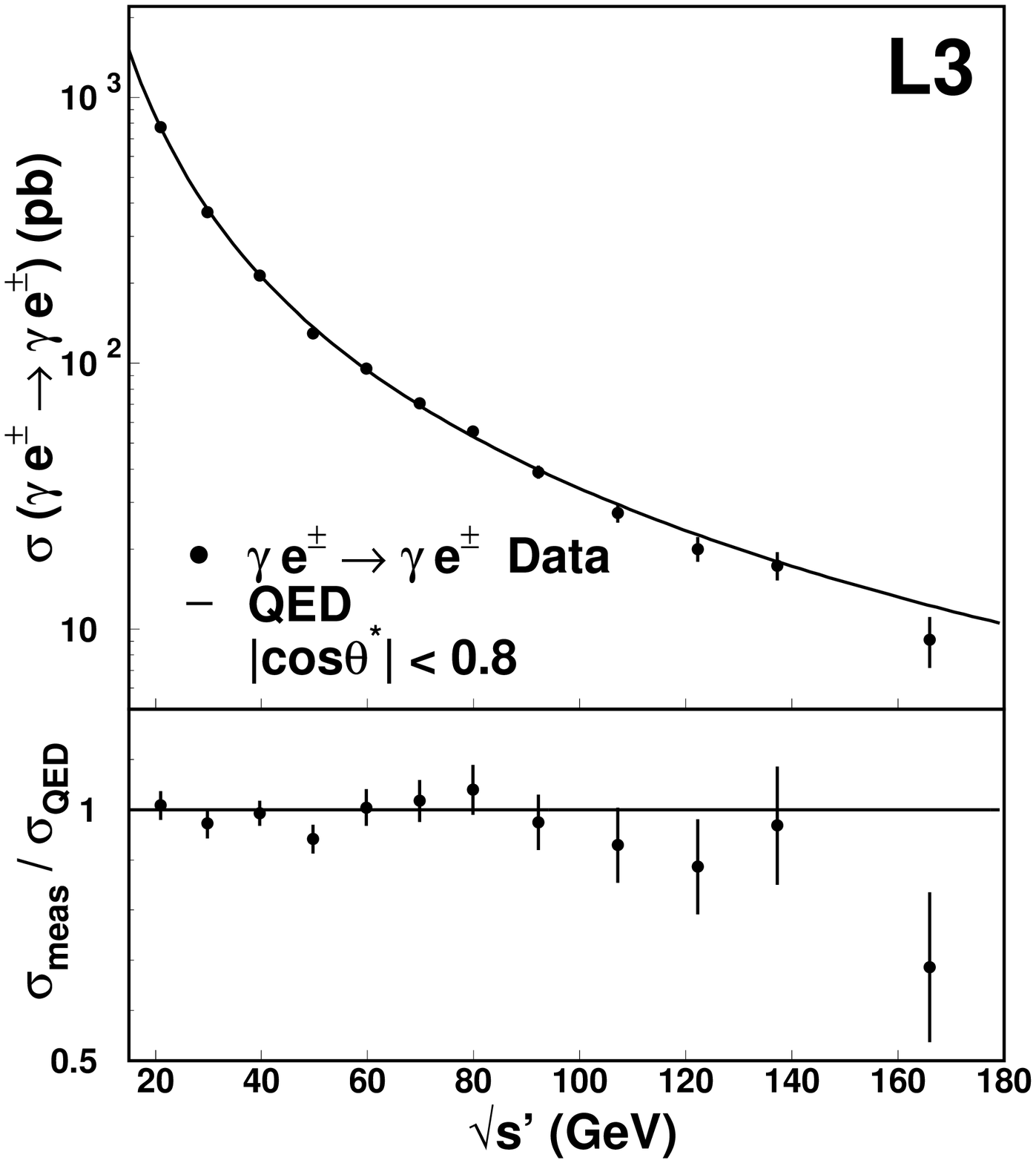,width=0.9\textwidth}}
    \icaption[]{\label{fig:7} Cross section of the $\rm\gamma
    e^{\pm}\ra\gamma e^{\pm}$ process measured as a function of the
    effective centre-of-mass energy, $\sqrt{s'}$, for
    $|\cos\theta^*|<0.8$, compared to the QED predictions. The full
    data sample collected at $\sqrt{s}=91.2-209.2\GeV$ is considered
    and the cosine of the electron rest-frame scattering angle is
    limited to the range $|\cos\theta^*|<0.8$.}
  \end{center}
\end{figure}

The cross sections as a function of $\sqrt{s'}$ measured in the range
$\sqrt{s'}=35-175\GeV$ are combined with those L3 measured in the
range $\sqrt{s'}=20-100\GeV$~\cite{L3Old}.  The full data-sample
collected by the L3 detector at $\sqrt{s}=91-209\GeV$ is therefore
considered, covering a range $\sqrt{s'}=20-175\GeV$.  The results are
presented in Table~\ref{tab:5} and Figure~\ref{fig:7}. They are in
good agreement, over two orders of magnitude,  with the QED predictions.

%

\newpage

%
%

\newpage
\typeout{   }     
\typeout{Using author list for paper 287 -  }
\typeout{$Modified: Jul 15 2001 by smele $}
\typeout{!!!!  This should only be used with document option a4p!!!!}
\typeout{   }
%
%
%
%
%
%

\newcount\tutecount  \tutecount=0
\def\tutenum#1{\global\advance\tutecount by 1 \xdef#1{\the\tutecount}}
\def\tute#1{$^{#1}$}
\tutenum\aachen            
\tutenum\nikhef            
\tutenum\mich              
\tutenum\lapp              
\tutenum\basel             
\tutenum\lsu               
\tutenum\beijing           
\tutenum\bologna           
\tutenum\tata              
\tutenum\ne                
\tutenum\bucharest         
\tutenum\budapest          
\tutenum\mit               
\tutenum\panjab            
\tutenum\debrecen          
\tutenum\dublin            
\tutenum\florence          
\tutenum\cern              
\tutenum\wl                
\tutenum\geneva            
\tutenum\hamburg           
\tutenum\hefei             
\tutenum\lausanne          
\tutenum\lyon              
\tutenum\madrid            
\tutenum\florida           
\tutenum\milan             
\tutenum\moscow            
\tutenum\naples            
\tutenum\cyprus            
\tutenum\nymegen           
\tutenum\caltech           
\tutenum\perugia           
\tutenum\peters            
\tutenum\cmu               
\tutenum\potenza           
\tutenum\prince            
\tutenum\riverside         
\tutenum\rome              
\tutenum\salerno           
\tutenum\ucsd              
\tutenum\sofia             
\tutenum\korea             
\tutenum\taiwan            
\tutenum\tsinghua          
\tutenum\purdue            
\tutenum\psinst            
\tutenum\zeuthen           
\tutenum\eth               

{
\parskip=0pt
\noindent
{\bf The L3 Collaboration:}
\ifx\selectfont\undefined
 \baselineskip=10.8pt
 \baselineskip\baselinestretch\baselineskip
 \normalbaselineskip\baselineskip
 \ixpt
\else
 \fontsize{9}{10.8pt}\selectfont
\fi
\medskip
\tolerance=10000
\hbadness=5000
\raggedright
\hsize=162truemm\hoffset=0mm
\def\r{\rlap,}
\noindent

P.Achard\r\tute\geneva\ 
O.Adriani\r\tute{\florence}\ 
M.Aguilar-Benitez\r\tute\madrid\ 
J.Alcaraz\r\tute{\madrid}\ 
G.Alemanni\r\tute\lausanne\
J.Allaby\r\tute\cern\
A.Aloisio\r\tute\naples\ 
M.G.Alviggi\r\tute\naples\
H.Anderhub\r\tute\eth\ 
V.P.Andreev\r\tute{\lsu,\peters}\
% ...
\newpage
\begin{list}{A}{\itemsep=0pt plus 0pt minus 0pt\parsep=0pt plus 0pt minus 0pt
                \topsep=0pt plus 0pt minus 0pt}
\item[\aachen]
 III. Physikalisches Institut, RWTH, D-52056 Aachen, Germany$^{\S}$
\item[\nikhef] National Institute for High Energy Physics, NIKHEF, 
     and University of Amsterdam, NL-1009 DB Amsterdam, The Netherlands
\item[\mich] University of Michigan, Ann Arbor, MI 48109, USA
\item[\lapp] Laboratoire d'Annecy-le-Vieux de Physique des Particules, 
     LAPP,IN2P3-CNRS, BP 110, F-74941 Annecy-le-Vieux CEDEX, France
\item[\basel] Institute of Physics, University of Basel, CH-4056 Basel,
     Switzerland
\item[\lsu] Louisiana State University, Baton Rouge, LA 70803, USA
\item[\beijing] Institute of High Energy Physics, IHEP, 
  100039 Beijing, China$^{\triangle}$ 
\item[\bologna] University of Bologna and INFN-Sezione di Bologna, 
     I-40126 Bologna, Italy
\item[\tata] Tata Institute of Fundamental Research, Mumbai (Bombay) 400 005, India
\item[\ne] Northeastern University, Boston, MA 02115, USA
\item[\bucharest] Institute of Atomic Physics and University of Bucharest,
     R-76900 Bucharest, Romania
\item[\budapest] Central Research Institute for Physics of the 
     Hungarian Academy of Sciences, H-1525 Budapest 114, Hungary$^{\ddag}$
\item[\mit] Massachusetts Institute of Technology, Cambridge, MA 02139, USA
\item[\panjab] Panjab University, Chandigarh 160 014, India
\item[\debrecen] KLTE-ATOMKI, H-4010 Debrecen, Hungary$^\P$
\item[\dublin] Department of Experimental Physics,
  University College Dublin, Belfield, Dublin 4, Ireland
\item[\florence] INFN Sezione di Firenze and University of Florence, 
     I-50125 Florence, Italy
\item[\cern] European Laboratory for Particle Physics, CERN, 
     CH-1211 Geneva 23, Switzerland
\item[\wl] World Laboratory, FBLJA  Project, CH-1211 Geneva 23, Switzerland
\item[\geneva] University of Geneva, CH-1211 Geneva 4, Switzerland
\item[\hamburg] University of Hamburg, D-22761 Hamburg, Germany
\item[\hefei] Chinese University of Science and Technology, USTC,
      Hefei, Anhui 230 029, China$^{\triangle}$
\item[\lausanne] University of Lausanne, CH-1015 Lausanne, Switzerland
\item[\lyon] Institut de Physique Nucl\'eaire de Lyon, 
     IN2P3-CNRS,Universit\'e Claude Bernard, 
     F-69622 Villeurbanne, France
\item[\madrid] Centro de Investigaciones Energ{\'e}ticas, 
     Medioambientales y Tecnol\'ogicas, CIEMAT, E-28040 Madrid,
     Spain${\flat}$ 
\item[\florida] Florida Institute of Technology, Melbourne, FL 32901, USA
\item[\milan] INFN-Sezione di Milano, I-20133 Milan, Italy
\item[\moscow] Institute of Theoretical and Experimental Physics, ITEP, 
     Moscow, Russia
\item[\naples] INFN-Sezione di Napoli and University of Naples, 
     I-80125 Naples, Italy
\item[\cyprus] Department of Physics, University of Cyprus,
     Nicosia, Cyprus
\item[\nymegen] Radboud University and NIKHEF, 
     NL-6525 ED Nijmegen, The Netherlands
\item[\caltech] California Institute of Technology, Pasadena, CA 91125, USA
\item[\perugia] INFN-Sezione di Perugia and Universit\`a Degli 
     Studi di Perugia, I-06100 Perugia, Italy   
\item[\peters] Nuclear Physics Institute, St. Petersburg, Russia
\item[\cmu] Carnegie Mellon University, Pittsburgh, PA 15213, USA
\item[\potenza] INFN-Sezione di Napoli and University of Potenza, 
     I-85100 Potenza, Italy
\item[\prince] Princeton University, Princeton, NJ 08544, USA
\item[\riverside] University of Californa, Riverside, CA 92521, USA
\item[\rome] INFN-Sezione di Roma and University of Rome, ``La Sapienza",
     I-00185 Rome, Italy
\item[\salerno] University and INFN, Salerno, I-84100 Salerno, Italy
\item[\ucsd] University of California, San Diego, CA 92093, USA
\item[\sofia] Bulgarian Academy of Sciences, Central Lab.~of 
     Mechatronics and Instrumentation, BU-1113 Sofia, Bulgaria
\item[\korea]  The Center for High Energy Physics, 
     Kyungpook National University, 702-701 Taegu, Republic of Korea
\item[\taiwan] National Central University, Chung-Li, Taiwan, China
\item[\tsinghua] Department of Physics, National Tsing Hua University,
      Taiwan, China
\item[\purdue] Purdue University, West Lafayette, IN 47907, USA
\item[\psinst] Paul Scherrer Institut, PSI, CH-5232 Villigen, Switzerland
\item[\zeuthen] DESY, D-15738 Zeuthen, Germany
\item[\eth] Eidgen\"ossische Technische Hochschule, ETH Z\"urich,
     CH-8093 Z\"urich, Switzerland
\item[\S]  Supported by the German Bundesministerium 
        f\"ur Bildung, Wissenschaft, Forschung und Technologie.
\item[\ddag] Supported by the Hungarian OTKA fund under contract
numbers T019181, F023259 and T037350.
\item[\P] Also supported by the Hungarian OTKA fund under contract
  number T026178.
\item[$\flat$] Supported also by the Comisi\'on Interministerial de Ciencia y 
        Tecnolog{\'\i}a.
\item[$\sharp$] Also supported by CONICET and Universidad Nacional de La Plata,
        CC 67, 1900 La Plata, Argentina.
\item[$\triangle$] Supported by the National Natural Science
  Foundation of China.
\end{list}
}
\vfill


\newpage

%
%

\begin{table}[htb]
  \begin{center}
    \begin{tabular}{|c|c|c|c|c|c|c|c|}
      \hline
      $\sqrt{s}$ (Ge\kern -0.1em V)  & 188.6 &  191.6 & 195.6 &  199.5 &  201.7 & $202.5-205.5$ &  $205.5-209.2$ \\   
      $\cal{L}$ (pb$^{-1})$ & 176.8 &   28.9 &  82.9 &   67.8 &   36.2 &    75.6 & 137.7 \\ 
      $N_{\rm data}$        &  1409 &   221  &  612  &   460  &   266  &    545  &   974 \\ 
      $N_{\rm MC}$          &  1336 &   231  &  659  &   521  &   284  &    520  &  982 \\ 
      \hline
    \end{tabular}
  \icaption[]{\label{tab:1}
    Integrated luminosities, $\cal{L}$, and numbers of events selected in data, $N_{\rm data}$,
    and Monte Carlo, $N_{\rm MC}$, at each $\epem$ centre-of-mass energy.}
  \end{center}
\end{table}

\begin{table}[htb]
  \begin{center}
    \begin{tabular}{|l|c|c|}
      \cline{2-3}
      \multicolumn{1}{c|}{} & \multicolumn{2}{c|}{Systematic uncertainty on}\\
      \cline{1-1}
      Source  & $\sigma_{\rm\gamma e^\pm \ra \gamma e^\pm}(\sqrt{s'})$ &  $\rm d \sigma_{\rm\gamma e^\pm \ra \gamma e^\pm}/d \cos{\theta^*}$\\
      \hline
      Measurement of angles              & $0.2 \% - 3.8 \%$ &  $0.1 \%-18.6 \%$ \\
      Energy scale                       & $0.1 \% - 0.9 \%$ &  $0.1 \%-\phantom{0}1.9 \%$ \\
      Charge confusion                   & $0.1 \% - 0.2 \%$ &  $0.1 \%-19.8 \%$ \\
      Background normalisation           & $0.1 \% - 0.2 \%$ &  $0.1 \%-\phantom{0}0.3 \%$ \\
      Signal Monte Carlo statistics      & $1.0 \% - 3.7 \%$ &  $1.3 \%-\phantom{0}6.9 \%$ \\
      Background Monte Carlo statistics  & $0.3 \% - 2.4 \%$ &  $0.3 \%-\phantom{0}4.2 \%$ \\ \hline
      Total                              & $1.3 \% - 5.5 \%$ &  $1.5 \%-28.2 \%$ \\ \hline
    \end{tabular}
  \icaption[]{\label{tab:2}
    Systematic uncertainties on the cross section as a function of
    $\sqrt{s'}$ and of the differential cross section as a function of
    $\cos\theta^*$. The largest uncertainties correspond to the
    low-statistics high $\sqrt{s'}$ regions and to the
    forward-scattering regions, where angular measurements and charge
    confusion effects are important.}
  \end{center}
\end{table}

\begin{table}[b]
  \begin{center}
    \begin{tabular}{|c|c|c|c|c|c|c|}
      \hline
      \rule {0pt}{12pt} $\sqrt{s'}$ (\rm Ge\kern -0.1em V)      &
      $\langle \sqrt{s'} \rangle$ (\rm Ge\kern -0.1em V)        & 
      $N_{\rm data}$                                            & 
      $N_{\rm MC}^{\rm sign}$                                   & 
      $N_{\rm MC}^{\rm back}$                                   & 
      $\sigma_{\rm\gamma e^\pm \ra \gamma e^\pm}$ (pb)          & 
      $\sigma_{\rm\gamma e^\pm \ra \gamma e^\pm}^{\rm QED}$ (pb)\\
      \hline
      $\phantom{0}35-\phantom{0}45$ &  \phantom{0}39.7 &  1269 & 1229.7 & 32.0 & $ 216.0 \pm  6.0 \pm 4.7 $  & $ 214.7$ \\
      $\phantom{0}45-\phantom{0}55$ &  \phantom{0}49.7 &  \phantom{0}889 &  \phantom{0}900.3 & 22.5 & $ 131.5 \pm  4.6 \pm 1.8 $  & $ 136.7$ \\
      $\phantom{0}55-\phantom{0}65$ &  \phantom{0}59.8 &  \phantom{0}627 &  \phantom{0}610.6 & 11.8 & $  \phantom{0}95.3 \pm  3.8 \pm 2.2 $  & $  \phantom{0}94.6$ \\
      $\phantom{0}65-\phantom{0}75$ &  \phantom{0}69.8 &  \phantom{0}506 &  \phantom{0}472.5 & \phantom{0}9.8 & $  \phantom{0}72.9 \pm  3.1 \pm 1.5 $  & $  \phantom{0}69.4$ \\
      $\phantom{0}75-\phantom{0}85$ &  \phantom{0}79.8 &  \phantom{0}370 &  \phantom{0}342.4 & 16.5 & $  \phantom{0}54.8 \pm  2.8 \pm 1.2 $  & $  \phantom{0}53.1$ \\
      $\phantom{0}85-100$ &  \phantom{0}92.2 &  \phantom{0}357 &  \phantom{0}346.9 & 14.9 & $  \phantom{0}39.2 \pm  2.1 \pm 1.1 $  & $  \phantom{0}39.8$ \\
      $100-115$ & 107.2 & \phantom{0}205 &  \phantom{0}214.4 &  \phantom{0}5.8 & $  \phantom{0}27.3 \pm  2.1 \pm 0.8 $  & $  \phantom{0}29.4$ \\
      $115-130$ & 122.3 & \phantom{0}125 &  \phantom{0}138.2 &  \phantom{0}2.6 & $  \phantom{0}20.0 \pm  2.0 \pm 0.7 $  & $  \phantom{0}22.6$ \\
      $130-145$ & 137.3 & \phantom{00}87 &  \phantom{00}86.6 &  \phantom{0}3.3 & $  \phantom{0}17.3 \pm  1.9 \pm 0.9 $  & $  \phantom{0}17.9$ \\
      $145-175$ & 159.3 & \phantom{00}52 &  \phantom{00}66.6 &  \phantom{0}6.5 & $  \phantom{00}9.1 \pm  1.8 \pm 0.7 $  & $  \phantom{0}13.3$ \\
      \hline
    \end{tabular}
    \icaption[]{\label{tab:3}
      Number of events observed in each $\sqrt{s'}$ bin with average
      $\langle\sqrt{s'}\rangle$, $N_{\rm data}$, together with the signal,
      $N_{\rm MC}^{\rm sign}$, and background, $N_{\rm MC}^{\rm back}$,
      Monte Carlo predictions.  The measured cross sections,
      $\sigma_{\rm\gamma e^\pm \ra \gamma e^\pm}$, are given with their
      statistical and systematic uncertainties, respectively, together with
      the QED predictions, $\sigma_{\rm\gamma e^\pm \ra \gamma
      e^\pm}^{\rm QED}$. The data sample at $\sqrt{s}=188.6-209.2\GeV$ is considered,
      and the cosine of the electron rest-frame scattering angle is limited
      to the range $|\cos\theta^*|<0.8$. }
  \end{center}
\end{table}

\begin{table}[ht]
  \begin{center}
    \begin{tabular}{|c|c|c|c|c|c|c|}
      \hline
      \rule {0pt}{12pt} $ \cos\theta^* $                                       &
      $\langle \cos\theta^* \rangle $                                          &
      $N_{\rm data}$                                                           & 
      $N_{\rm MC}^{\rm sign}$                                                  & 
      $N_{\rm MC}^{\rm back}$                                                  & 
      $\rm d\sigma_{\rm\gamma e^\pm \ra \gamma e^\pm}/ d \cos\theta^*$ (pb)     & 
      $\rm d\sigma_{\rm\gamma e^\pm \ra \gamma e^\pm}^{\rm QED}/d \cos\theta^*$ (pb) \\
      \hline
      $-0.80 - -0.67$ & $ -0.74$ & 346 & 283.3 & 25.8 & $ 133.8 \pm  6.4 \pm 9.6  \times 10^{2}$  & $ 118.3 \times 10^{2}$ \\
      $-0.67 - -0.53$ & $ -0.60$ & 389 & 385.2 & 15.7 & $  \phantom{0}77.2 \pm  4.0 \pm 2.6 \times 10^{2}$  & $   \phantom{0}79.6\times 10^{2}$ \\
      $-0.53 - -0.40$ & $ -0.47$ & 499 & 469.9 & 11.2 & $  \phantom{0}63.6 \pm  2.7 \pm 2.5 \times 10^{2}$  & $   \phantom{0}61.2\times 10^{2}$ \\
      $-0.40 - -0.27$ & $ -0.34$ & 553 & 593.5 & 14.5 & $  \phantom{0}46.0 \pm  2.1 \pm 1.5 \times 10^{2}$  & $   \phantom{0}50.7\times 10^{2}$ \\
      $-0.27 - -0.13$ & $ -0.20$ & 742 & 733.8 & 10.1 & $  \phantom{0}43.1 \pm  1.6 \pm 0.7 \times 10^{2}$  & $   \phantom{0}44.1\times 10^{2}$ \\
      $-0.13 - \phantom{-}0.0\phantom{0}$  & $ -0.07$ & 616 & 650.2 &  \phantom{0}7.9 & $  \phantom{0}37.0 \pm  1.6 \pm  0.6\times 10^{2} $  & $  \phantom{0}39.6\times 10^{2}$ \\
      $\phantom{-}0.0\phantom{0} - \phantom{-}0.13$    & $ \phantom{-}0.06$ & 512 & 509.2 &  \phantom{0}5.8 & $  \phantom{0}36.4 \pm  1.6 \pm  0.6\times 10^{2} $  & $  \phantom{0}36.6\times 10^{2}$ \\
      $\phantom{-}0.13 - \phantom{-}0.27$   & $ \phantom{-}0.20$ & 353 & 347.6 & 10.9 & $  \phantom{0}33.4 \pm  1.8 \pm  0.8\times 10^{2} $  & $  \phantom{0}34.4\times 10^{2}$ \\
      $\phantom{-}0.27 - \phantom{-}0.40$   & $ \phantom{-}0.33$ & 221 & 202.4 &  \phantom{0}5.1 & $  \phantom{0}35.0 \pm  2.2 \pm 2.2\times 10^{2} $  & $  \phantom{0}32.9\times 10^{2}$ \\
      $\phantom{-}0.40 - \phantom{-}0.53$   & $ \phantom{-}0.46$ & 135 & 126.7 & 11.9 & $  \phantom{0}30.1 \pm  2.7 \pm 3.1\times 10^{2} $  & $  \phantom{0}31.8\times 10^{2}$ \\
      $\phantom{-}0.53 - \phantom{-}0.67$   & $ \phantom{-}0.60$ &  \phantom{0}81 &   \phantom{0}73.5 & \phantom{0}2.8 & $  \phantom{0}33.1 \pm  3.4 \pm 5.9\times 10^{2} $  & $  \phantom{0}31.1\times 10^{2}$ \\
      $\phantom{-}0.67 - \phantom{-}0.80$   & $ \phantom{-}0.73$ &  \phantom{0}40 & \phantom{0}33.1 & \phantom{0}4.1 & $  \phantom{0}33.3 \pm  4.8 \pm 8.7\times 10^{2} $  & $  \phantom{0}30.7\times 10^{2}$ \\
      \hline
    \end{tabular}
    \icaption[]{\label{tab:4}
      Number of events observed in each $\cos\theta^*$ bin with average
      $\langle \cos\theta^* \rangle$, $N_{\rm data}$, together with the signal, $N_{\rm
	MC}^{\rm sign}$, and background, $N_{\rm MC}^{\rm back}$, Monte Carlo
      predictions. The measured differential cross sections, $\rm
      d\sigma_{\rm\gamma e^\pm \ra \gamma e^\pm}/ d \cos\theta^*$, are given
      with their statistical and systematic uncertainties, respectively,
      together with the QED predictions, $\rm d\sigma_{\rm\gamma e^\pm \ra
	\gamma e^\pm}^{\rm QED}/ d \cos\theta^*$. The data sample at
      $\sqrt{s}=188.6-209.2\GeV$ is considered, corresponding to an
      effective centre-of-mass energy range $\sqrt{s'}=35-175\GeV$.}
  \end{center}
\end{table}

\begin{table}[hb]
  \begin{center}
    \begin{tabular}{|c|c|c|c|}
      \hline
      \rule {0pt}{12pt} $\sqrt{s'} (\rm Ge\kern -0.1em V)$ &
      $\langle \sqrt{s'} \rangle (\rm Ge\kern -0.1em V)$   &
      $\sigma_{\rm\gamma e^\pm \ra \gamma e^\pm}$          &
      $\sigma_{\rm\gamma e^\pm \ra \gamma e^\pm}^{\rm QED}$\\
      \hline
      $\phantom{00}0-\phantom{0}25 $   & $\phantom{0}21.0$  & $771.2 \pm 21.6$ & 764.8 \\
      $\phantom{0}25-\phantom{0}35$   & $\phantom{0}29.8$  & $370.6 \pm 11.3$ & 381.1 \\
      $\phantom{0}35-\phantom{0}45$   & $\phantom{0}39.7$  & $213.2 \pm  \phantom{0}5.4$ & 214.7 \\
      $\phantom{0}45-\phantom{0}55$   & $\phantom{0}49.7$  & $128.7 \pm  \phantom{0}3.9$ & 136.7 \\
      $\phantom{0}55-\phantom{0}65$   & $\phantom{0}59.8$  & $ \phantom{0}95.0 \pm  \phantom{0}3.5$ &  \phantom{0}94.6 \\
      $\phantom{0}65-\phantom{0}75$   & $\phantom{0}69.8$  & $ \phantom{0}70.6 \pm  \phantom{0}2.9$ &  \phantom{0}69.4 \\
      $\phantom{0}75-\phantom{0}85$   & $\phantom{0}79.8$  & $ \phantom{0}55.2 \pm  \phantom{0}2.6$ &  \phantom{0}53.1 \\
      $\phantom{0}85-100$  &  $\phantom{0}92.2$ & $ \phantom{0}38.8 \pm  \phantom{0}2.2$ &  \phantom{0}39.8 \\
      $100-115$ & $107.2$ & $ \phantom{0}27.3 \pm  \phantom{0}2.2$ &  \phantom{0}29.4 \\
      $115-130$ & $122.3$ & $ \phantom{0}20.0 \pm  \phantom{0}2.1$ &  \phantom{0}22.6 \\
      $130-145$ & $137.3$ & $ \phantom{0}17.3 \pm  \phantom{0}2.1$ &  \phantom{0}17.9 \\
      $145-175$ & $159.3$ & $  \phantom{00}9.1 \pm  \phantom{0}2.0$ &  \phantom{0}13.3 \\
      \hline
    \end{tabular}
    \icaption[]{\label{tab:5} Measured cross sections,
      $\sigma_{\rm\gamma e^\pm \ra \gamma e^\pm}$, with their combined
      statistical and systematic uncertainties together with the QED
      predictions, $\sigma_{\rm\gamma e^\pm \ra \gamma e^\pm}^{\rm
      QED}$. The full L3 data sample at $\sqrt{s}=91.2-209.2\GeV$ is
      considered and the cosine of the electron rest-frame scattering
      angle is limited to the range $|\cos\theta^*|<0.8$.}
  \end{center}
\end{table}
\clearpage

%
%

\newpage
               
\begin{figure}[p]
  \begin{center}
    \begin{tabular}{c}
      \mbox{\epsfig{figure=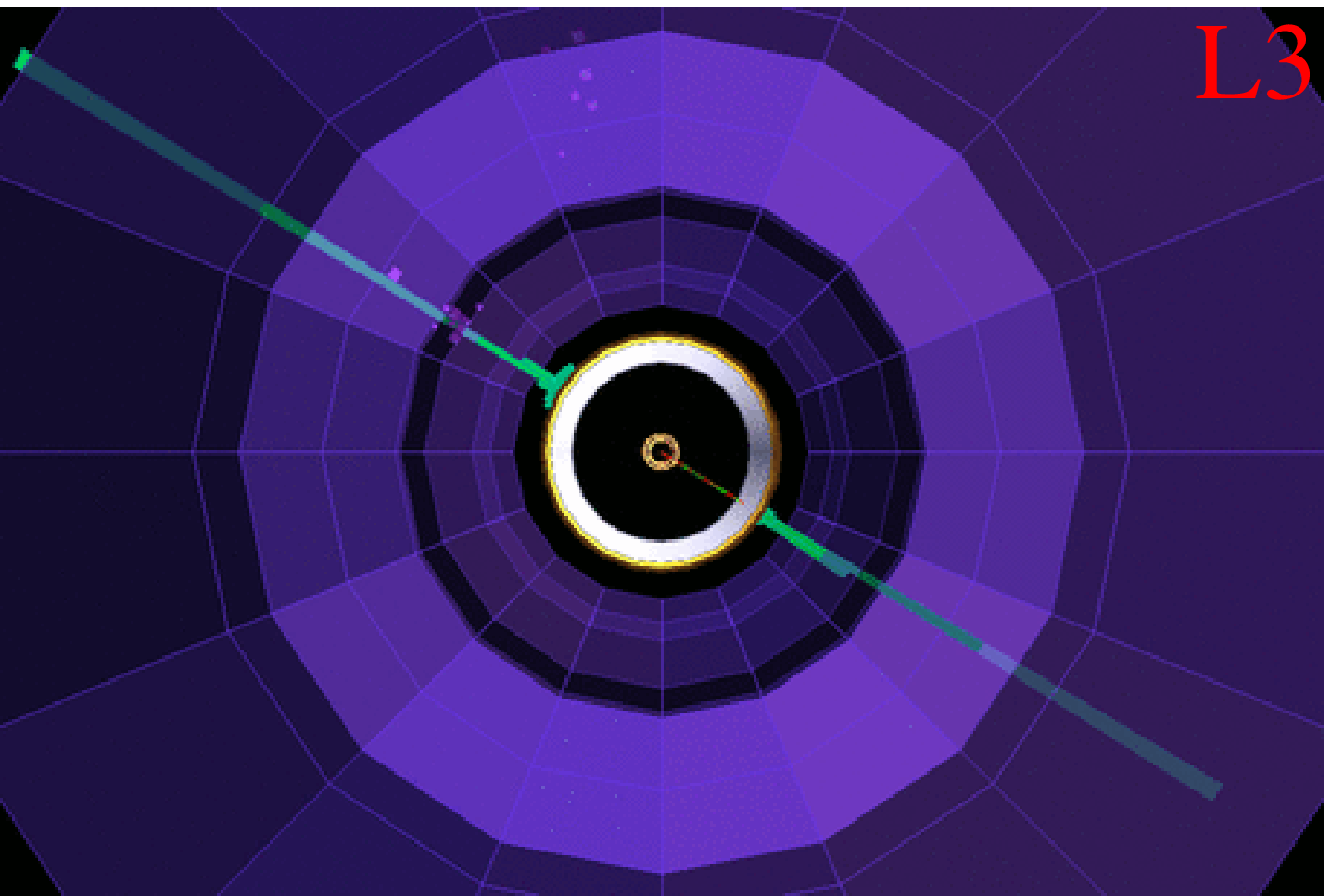,width=0.8\textwidth}}\\
      \mbox{\epsfig{figure=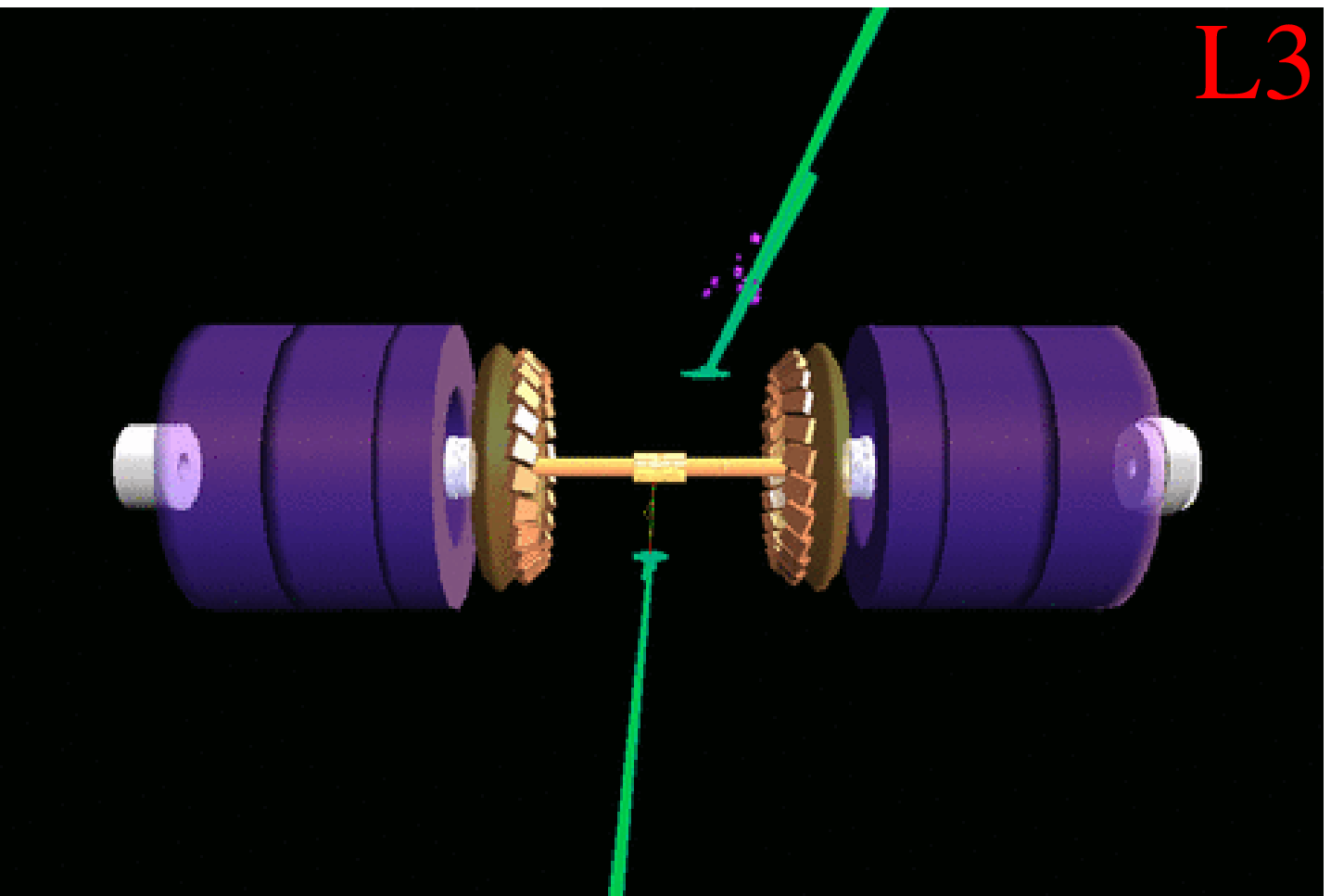,width=0.8\textwidth}}\\
    \end{tabular}
    \icaption[]{\label{fig:5} Views of a high-energy
    candidate for quasi-real Compton scattering in the plane
    transverse to the beams, higher plot, and in a plane along the
    beams, lower plot. The towers represent the energy deposited in
    the electromagnetic calorimeter and the boxes are low-energy clusters
    in the hadron calorimeter. The track of the electron is clearly
    visible, while no track is associated to the other electromagnetic
    cluster, identified as the photon. }
  \end{center}
\end{figure}

\end{document}